\documentclass[pre,aps,floats,twocolumn,superscriptaddress,floatfix]{revtex4}
\usepackage{graphicx,amssymb,amsmath,ifthen,rotating}
\usepackage[active]{srcltx}
\usepackage{color}
\begin{document}
\title{Monodisperse hard rods in external potentials}  
\author{Benaoumeur Bakhti}
\affiliation{
  Fachbereich Physik,
  Universit\"at Osnabr\"uck,
  D-49076 Osnabr\"uck, Germany}
    \author{Michael Karbach}
\affiliation{
  Fachbereich Physik,
  Bergische Universit\"at Wuppertal,
  D-42097 Wuppertal, Germany}
    \author{Philipp Maass}
\affiliation{
  Fachbereich Physik,
  Universit\"at Osnabr\"uck,
  D-49076 Osnabr\"uck, Germany}
\author{Gerhard M{\"{u}}ller}
\affiliation{
  Department of Physics,
  University of Rhode Island,
  Kingston RI 02881, USA}

\begin{abstract}
We consider linear arrays of cells of volume $V_\mathrm{c}$ populated by monodisperse rods of size $\sigma V_\mathrm{c}$, $\sigma=1,2,\ldots$, subject to hardcore exclusion interaction.
Each rod experiences a position-dependent external potential.
In one application we also examine effects of contact forces between rods.
We employ two distinct methods of exact analysis with complementary strengths and different limits of spatial resolution to calculate profiles of pressure and density on mesoscopic and microscopic length scales at thermal equilibrium.
One method uses density functionals and the other statistically interacting vacancy particles.
The applications worked out include gravity, power-law traps, and hard walls.
We identify oscillations in the profiles on a microscopic length scale and show how they are systematically averaged out on a well-defined mesoscopic length scale to establish full consistency between the two approaches.
The continuum limit, realized as $V_\mathrm{c}\to0$, $\sigma\to\infty$ at nonzero and finite $\sigma V_\mathrm{c}$, connects our highest-resolution results with known exact results for monodisperse rods in a continuum.
We also compare the pressure profiles obtained from density functionals with the average microscopic pressure profiles derived from the pair distribution function.

\end{abstract}

\maketitle

%
\section{Introduction}\label{sec:intro}
%
In classical statistical mechanics, particles with shapes are ubiquitous.
Their prominence in granular matter \cite{Capriz/etal:2008,Mehta:1994}, soft condensed matter \cite{Jones:2002,Doi:2012}, and, more specifically, biological matter \cite{Bialeki:2012,Nelson:2013}, is well established.
Their shapes vary from the highly complex such as folded proteins to the most elementary such as hard spheres.
Their equilibrium and nonequilibrium properties are investigated by a broad array of experimental and computational probes.

Analytic approaches in this area of research see their predictive power restricted to fairly simple scenarios regarding shapes, environment, and dimensionality.  
For rigorous calculations the domains of applicability are further narrowed. 
This limitation finds ample compensation in their usefulness as benchmarks and anchor points for approximations and simulations. 
The work reported in the following is motivated by this chain of reasoning. 
It deals with hard rods in one dimension at thermal equilibrium in external potentials.  
We consider monodisperse rods of size $\sigma V_\mathrm{c}$, $\sigma=1,2,\ldots$ on a lattice (linear array of cells with volume $V_\mathrm{c}$). 
Hard rods of size $V_\mathrm{r}$ populating a continuum emerge from the limit $V_\mathrm{c}\to0$, $\sigma\to\infty$ with $\sigma V_\mathrm{c}=V_\mathrm{r}$.

There are several approaches that facilitate an exact derivation of thermodynamic and structural properties for a homogeneous one-dimensional hard rod system (Tonks gas) with first-neighbor Takahashi-type interactions \cite{Thompson:1972}. 
One rather elegant method uses convolution relations between Boltzmann factors to determine partition functions \cite{Bishop/Boonstra:1983}. 
It produces the free enthalpy and the equation of state (EOS) for first-neighbor interactions of arbitrary range.
Many-body-density distribution functions can be calculated by an extension of this approach \cite{Davis:1990}. 
An alternative method of similar scope has recently been developed.
It uses statistically interacting vacancy particles (SIVP) as quasiparticles \cite{Bakhti/etal:2014}. 
This method also yields the size distribution of vacancies between rods.

On the basis of these approaches for homogeneous systems, it is possible to treat inhomogeneous systems by assuming that the EOS is valid on a coarse-grained local scale. 
From the requirement of mechanical equilibrium, spatial variations of pressure and density can then be calculated from the spatial variations of the external potential. 
This provides a simple and common thermodynamic route for calculating density and pressure profiles \cite{Landau/Lifshitz:1986}, which generally is a difficult task for an interacting many-particle system. 
We will refer to this thermodynamic route as the EOS method in the following. 
Because the EOS method relies on the assumption of the existence of a local EOS, it is interesting to gain insight into how far this assumption is justified and whether the method gives useful information even if the underlying assumption does not hold.

To tackle this question analytically, exact results for inhomogeneous systems are required. 
For hard rods with first-neighbor interactions in one dimension, exact treatments are possible via recursion relations for partition functions \cite{Davis:1990, Buschle/etal:2000a} or density functional theory (DFT) \cite{Percus:1982,Buschle/etal:2000b,Bakhti/etal:2012}. 
These methods allow for the exact derivation of density profiles as well as many-body distribution functions.
Given exact density profiles, pressure profiles can be obtained by resorting to the 
requirement of mechanical
equilibrium as in the EOS method, but without assuming a local EOS.

The exact calculation of local pressures is a more subtle task. 
Generally, the local pressure can be defined via the trace of the local stress tensor, which governs the time evolution of the momentum density in a coarse-grained continuum description \cite{Irving/Kirkwood:1950, Lutsko:1988}. 
In case of pair interaction forces, thermodynamic averaging over the corresponding local stress tensor allows for the determination of pressure profiles from the density profiles and pair distribution function.

Our calculations here will employ two approaches. 
The DFT for lattice fluids \cite{Nieswand/etal:1993a, Reinel/Dieterich:1996, Reinhard/etal:2000, Buschle/etal:2000a, 
Buschle/etal:2000b, Bakhti/etal:2013} is used to determine exact density profiles in external fields and the SIVP approach \cite{Haldane:1991, Wu:1994, Isakov:1994, Anghel:2007,   Liu/etal:2011, Liu/etal:2012, Bakhti/etal:2014} is used as a realization of the EOS method. 
The DFT and SIVP approaches have domains that partially overlap and strengths that complement each other. 
In the DFT, the operational degrees of freedom are the rods themselves. 
In the SIVP approach, the operational degrees of freedom are the vacancies between the rods. 

We begin by describing the general methodology and background (Sec.~\ref{sec:meth}) and then proceed with applications to rods in a uniform gravitational field (Sec.~\ref{sec:grav}) and in a power-law trap (Sec.~\ref{sec:powe}). 
The subtleties regarding average microscopic pressure are addressed in the context of the first application.
Steric wall effects in lattice systems and their relations to known continuum results are discussed in Sec.~\ref{sec:ster}.
In Sec.~\ref{sec:conc} we summarize the main conclusions and outline projected extensions to polydisperse rods.
Appendices \ref{sec:appa} and \ref{sec:appc} summarize outlying background materials for use in the main text.
Appendix \ref{sec:appb} presents a highly practical method of calculating exact density profiles within the DFT framework for arbitrary external potentials.

%
\section{Methodology}\label{sec:meth}
%

\subsection{Model system}\label{sec:meth-mode}
\begin{figure}[b]
  \begin{center}
 \includegraphics[width=85mm]{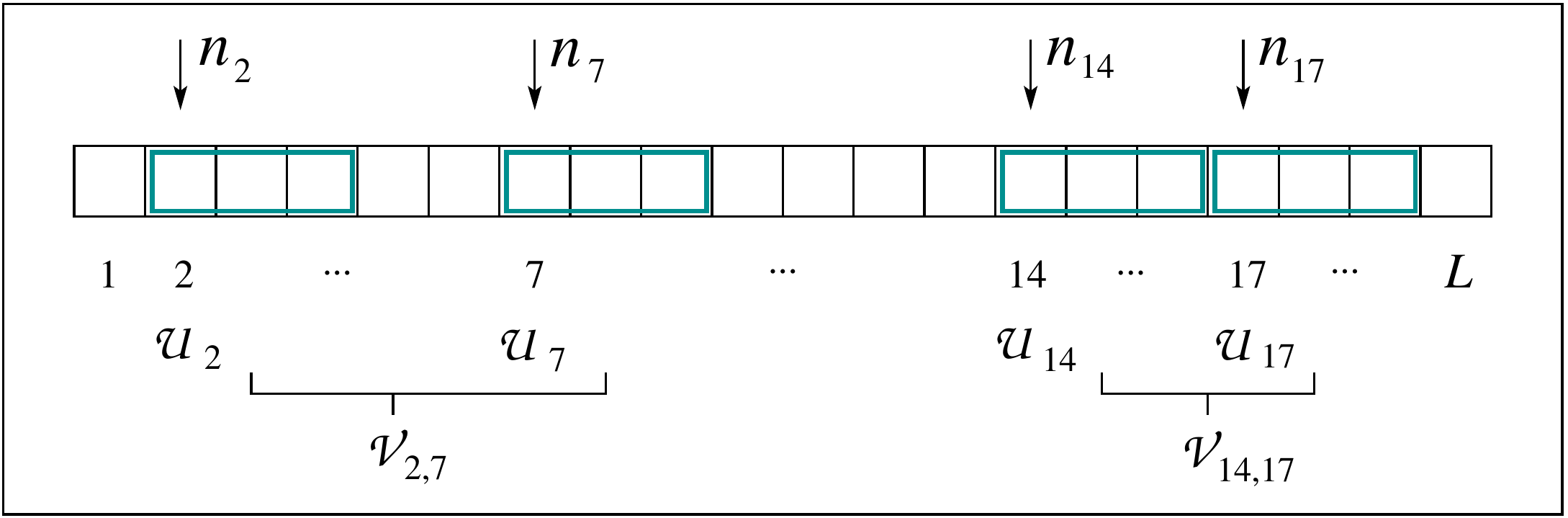}
\end{center}
\caption{Four rods of size $\sigma=3$ at positions $i=2,7,14,17$ on a
  lattice of size $L=20$.  The relevant external potentials
  $\mathcal{U}_i$ and interaction potentials $\mathcal{V}_{i,j}$ are
  stated.  One is a contact potential and the other represents an
  interaction of maximum range $(\sigma-1)V_\mathrm{c}$.}
  \label{fig:sketch-dft}
\end{figure}

Consider rods on a linear chain $i=1,\ldots,L$ of lattice sites, represented in Fig.~\ref{fig:sketch-dft} as an array of cells.  
Each cell has volume $V_\mathrm{c}$.
Rods of size $\sigma$ occupy that many adjacent cells. 
Hard walls at both ends of the chain define the boundary conditions.  
Assuming that cells (and rods) have unit cross section we can conveniently use $V_\mathrm{c}$ as a microscopic scale for both volume and length.

Microstates of this system are encoded in a sequence of occupation numbers, $\mathbf{n}\doteq\{n_1,\ldots,n_L\}$, $n_i=0,1$.
To a rod that occupies sites $i,\ldots,i+\sigma-1$, we assign the occupation number $n_i=1$.  
Hardcore exclusion imposes the conditions $n_in_{i+j}=0$, $j=1,\ldots,\sigma-1$, and the hard walls imply $n_i=0$ for $i<1$ and $i>L-\sigma+1$, respectively.

The model system analyzed in this work is specified by the Hamiltonian,
\begin{equation}\label{eq:hamiltonian} 
 \mathcal{H}(\mathbf{n})=
\sum_{i<j}\mathcal{V}_{i,j}n_in_j+\sum_{i}\mathcal{U}_in_i,
\end{equation}
where $\mathcal{U}_i$ is an external potential and $\mathcal{V}_{i,j}$ an interaction potential restricted to first-neighbor rods.
The first-neighbor restriction is naturally ascertained by interactions $\mathcal{V}_{i,j}$ with a range limited to $j=i+\xi$,
where $\sigma\leq\xi<2\sigma$, but an extended range between first-neighbor rods is permitted.

In this work we mainly examine the effects of hardcore repulsion in combination with external potentials.
Contact forces are included in one application (Secs.~\ref{sec:grav-sivp-cont} and \ref{sec:grav-dft-cont}).
The effects of long-range forces are being analyzed in a separate study \cite{selgra}.

\subsection{Exact density functionals}\label{sec:meth-exac}
The analysis carried out in \cite{Bakhti/etal:2013} is based on former work \cite{Nieswand/etal:1993a,   Reinel/Dieterich:1996, Reinhard/etal:2000, Buschle/etal:2000a, Buschle/etal:2000b} and expresses the grand potential as a density functional, i.e. a functional of the mean occupation numbers of rods $\tilde{n}_i\doteq\langle n_i\rangle$ 
\footnote{The averages are over the (grand-canonical) Boltzmann distribution for Hamiltonian (\ref{eq:hamiltonian}), but with $\mathcal{U}_i$ replaced by $\mathcal{U}_i^{\scriptscriptstyle\rm Mermin}$, where $\mathcal{U}_i^{\scriptscriptstyle\rm Mermin}$ is the unique external  `Mermin potential' that (for given interaction $\mathcal{V}_{i,j}$) would generate the $\bar n_i$ as the mean occupation numbers in equilibrium.}
\begin{equation}\label{eq:omega}
\Omega[\tilde n_1,\ldots,\tilde n_L]=F[\tilde n_1,\ldots,\tilde n_L]+
\sum_{i=1}^{L}(\mathcal{U}_i-\mu)\tilde n_i\,,
\end{equation}
where $\mu$ is the chemical potential. 
The intrinsic free-energy functional in Eq.~(\ref{eq:omega}) can be written in the form 
\begin{align}\label{eq:freeenergy-dft}
F[\tilde{\mathbf{n}}]&=\sum_{i=1}^L f_i[\tilde{\mathbf{n}}]
=\sum_{i=1}^L (e_i[\tilde{\mathbf{n}}]-Ts_i[\tilde{\mathbf{n}}])\,,
\end{align}
where $T$ is the temperature, and $f_i[\tilde{\mathbf{n}}]=e_i[\tilde{\mathbf{n}}]-Ts_i[\tilde{\mathbf{n}}]$, $e_i[\tilde{\mathbf{n}}]$, and $s_i[\tilde{\mathbf{n}}]$ are local functionals of the intrinsic free energy, internal interaction energy, and entropy. 
The latter are given by
\begin{align}\label{eq:internalenergy-dft}
e_i[\tilde{\mathbf{n}}]=\sum_{j=i+\sigma}^{i+\xi}
\mathcal{V}_{i,j}C_{i,j}[\tilde{\mathbf{n}}]
\end{align}
and 
\begin{align}\label{eq:entropy-dft}
s_i[\tilde{\mathbf{n}}]=&-k_\mathrm{B}\Bigl\{\Phi(a_i[\tilde{\mathbf{n}}])
+\Phi(b_{i,i}[\tilde{\mathbf{n}}])-\Phi(b_{i-1,i}[\tilde{\mathbf{n}}])
+\nonumber\\ &\hspace*{-2em}{}+
\sum_{j=i-\xi}^{i-\sigma}\bigl[\Phi(C_{i,j}[\tilde{\mathbf{n}}])
  +\Phi(d_{i,j}[\tilde{\mathbf{n}}])-
  \Phi(d_{i-1,j}[\tilde{\mathbf{n}}])\bigr] \Bigr\}
\end{align}
with pair correlators
\begin{align}
C_{i,j}[\tilde{\mathbf{n}}]=\langle n_in_j\rangle,
\label{eq:correlators}
\end{align}
$\Phi(x)\doteq x\ln x$, and 
\footnote{As shown in Ref.~\cite{Bakhti/etal:2013} the $a_i$, $b_{i,j}$ and $d_{i,j}$ refer to the probabilities to have zero or one rod in `one-particle cavities', i.e. in intervals, where at most one rod can be placed.}
\begin{subequations}
\label{eq:abc}
\begin{equation}
\label{eq:alpha}
a_i[\tilde{\mathbf{n}}]=\tilde n_i
-\sum_{j=i-\xi}^{i-\sigma}C_{i,j}[\tilde{\mathbf{n}}]\,,
\end{equation}
\begin{equation}
\label{eq:beta}
b_{i,j}[\tilde{\mathbf{n}}]=
1-\sum_{k=j-\xi}^i\tilde n_k
+\sum_{k=j-\xi+\sigma}^{i}\sum_{l=j-\xi}^{k-\sigma} C_{k,l}[\tilde{\mathbf{n}}]\,,
\end{equation}
\begin{equation}
\label{eq:gamma}
d_{i,j}[\tilde{\mathbf{n}}]=\tilde n_j
-\sum_{k=j+\sigma}^iC_{k,j}[\tilde{\mathbf{n}}]\,.
\end{equation}
\end{subequations}
The pair correlators (\ref{eq:correlators})
have their dependence on $\tilde{\mathbf{n}}$ encoded in the implicit relations,
\begin{equation}
\label{eq:corr-relation}
C_{i,j}=\frac{a_{i}d_{i,j}}{b_{i,i}}
\prod_{k=i+1}^{j+\xi}
\frac{d_{k,j}b_{k-1,k}}{d_{k-1,j}b_{k,k}}
e^{-\beta\mathcal{V}_{i,j}}\,,\quad \beta\doteq\frac{1}{k_\mathrm{B}T}.
\end{equation}

The equilibrium density profile of rods follows from the extremum condition,
\begin{equation}\label{eq:structure-equations} 
\frac{\partial}{\partial \tilde n_i}\Omega[\tilde{\mathbf{n}}]=0,
\quad i=1,\ldots,L.
\end{equation}
With the solution $\bar{\mathbf{n}}$ of (\ref{eq:structure-equations}), the functions $f_i[\bar{\mathbf{n}}]$, $e_i[\bar{\mathbf{n}}]$ and $s_i[\bar{\mathbf{n}}]$ become the intrinsic free energy, internal interaction energy, and entropy per site. 
The equilibrium density profile $\bar{\mathbf{n}}$ and the profiles of the thermodynamic potentials depend on temperature, crowding, interaction, and environment via $\beta$, $\mu$, $\mathcal{V}_{i,j}$, and $\mathcal{U}_i$, respectively.
The cell occupancy (mass density) is obtained from the rod occupancy (number density) via
\begin{equation}\label{eq:101} 
\rho_i\doteq\sum_{j=i-\sigma+1}^i \bar{n}_j.
\end{equation}

In homogeneous systems the pressure $p$ follows rigorously from the free-energy density $f\doteq F/LV_\mathrm{c}$ via 
$p=\bar{n}\,\mathrm{d}f/\mathrm{d}\bar{n}-f$.
A natural extension of this relation to systems with external potential has the form
\begin{align}
\label{eq:pressure-dft-direct2}
p_i=-f_i[\bar{\mathbf{n}}]+\sum_{j=i-\sigma+1}^i
\bar{n}_j\frac{\partial f_i[\bar{\mathbf{n}}]}{\partial \bar{n}_j}\,.
\end{align}
and produces pressure profiles on a microscopic length scale.
However, there is no guarantee that the pressure thus derived coincides with the average microscopic pressure as commonly defined via the pair-distribution function. 
More on this question follows in Secs.~\ref{sec:meth-loca} and \ref{sec:grav-micro}.

Exact profiles for $\bar{n}_i$ and $\rho_i$ for arbitrary external potentials and interactions limited to hardcore repulsion on the lattice are calculated by the method introduced in Appendix \ref{sec:appb}.

\subsection{Coarse graining}\label{sec:meth-coar}
For this comparative study of methods we need a continuum description for rod positions on a microscopic length scale used in the DFT approach (Sec.~\ref{sec:meth-exac}) that carries over naturally to the mesoscopic length scale used in any of the EOS methods, specifically the SIVP approach (Sec.~\ref{sec:meth-eosm}).
This continuum description of the lattice system is unrelated to the continuum limit.
We replace each lattice site $i$ with the interval ${[iV_\mathrm{c},(i+1)V_\mathrm{c}[}$ across 
one cell (of unit cross section), 
and we define the interaction potential $v(x,x')$ and external potential $u(x)$
by setting $v(x_i,x_j)=\mathcal{V}_{i,j}$ and $u(x_i)=\mathcal{U}_i$.
The first-neighbor restriction for the interaction potential becomes $v(x,x')=0$ for $|x-x'|\geq\sigma V_\mathrm{c}$.
The local coverage at equilibrium of this interval allows us to define a number density  by the piecewise continuous function
\begin{equation}\label{eq:rho-dft} 
\rho(x)=\frac{\rho_i}{\sigma}=\frac{1}{\sigma}\sum_{j=i-\sigma+1}^{i}
\bar{n}_j,\quad x\in{[iV_\mathrm{c},(i+1)V_\mathrm{c}[}\,.
\end{equation}
For a homogeneous situation we have $\rho(x)=\bar{n}$.

From any density profile $\rho(x)$ we can calculate the associated pressure profiles $p(x)$ by invoking the balance between internal and external forces at thermodynamic equilibrium \cite{Landau/Lifshitz:1986},
\begin{equation}\label{eq:pressure-density-relation}
V_\mathrm{c}\frac{\mathrm{d}p(x)}{\mathrm{d}x}=\rho(x)f_\mathrm{u}(x)\,,
\end{equation}
where $f_\mathrm{u}=-\mathrm{d}u/\mathrm{d}x$ is the external force field \footnote{In one dimension, pressure has the same dimension as force, namely energy per length. The scaling factor $V_\mathrm{c}$ in (\ref{eq:pressure-density-relation}) makes $\rho(x)$ dimensionless, consistent with (\ref{eq:rho-dft}).}. 
Integration of this differential equation yields
\begin{equation}\label{eq:pressure-profile-dft}
p(x)-p(x_0)=\frac{1}{V_\mathrm{c}}\int_{x_0}^x \mathrm{d}x'\, \rho(x')f_\mathrm{u}(x')\,.
\end{equation}
In some applications the reference pressure $p(x_0)$ is known, e.g. via the weight of the rods in a uniform gravitational field.
In other cases it can be determined from the average number $N$ of rods, which we know from summing $\bar{n}_j$ over all sites, by using the normalization relation 
\begin{equation}\label{eq:301} 
 N=\frac{1}{V_\mathrm{c}}\int dx\,\rho(x)=\int dx\,\frac{p'(x)}{f_\mathrm{u}(x)},\quad 
p'\doteq\frac{dp}{dx}.
\end{equation}
If $x_0=LV_\mathrm{c}$ then we can use $p(x_0)=-\partial F[\bar{\mathbf{n}}]/\partial {x_0}$.
If the rods are only subject to hardcore repulsion we can use the fact that  at the system boundaries we have kinematic pressure $k_\mathrm{B}T\rho$ (Sec.~\ref{sec:meth-loca}).
In Sec.~\ref{sec:grav-dft} we compare profiles inferred from (\ref{eq:pressure-dft-direct2}) with profiles calculated from Eq.~(\ref{eq:pressure-profile-dft}).

\subsection{EOS method and SIVP approach}\label{sec:meth-eosm}
If one assumes that the EOS of a homogeneous system remains valid in a corresponding inhomogeneous system on a coarse-grained local scale in the presence of external potentials, the balance equation (\ref{eq:pressure-density-relation}) is sufficient to determine density and pressure profiles.
Depending on the circumstances we use (\ref{eq:pressure-density-relation}) to calculate the functions $p(x)$ and $\tilde{\rho}\big(p(x)\big)$ or the functions $\rho(x)$ and $\tilde{p}\big(\rho(x)\big)$
\footnote{In the case presence of phase transitions only the latter combination may be unique.}.

In the former case we have an EOS in the form $\tilde{\rho}(p)$ and solve (\ref{eq:pressure-density-relation}) by separating variables $p$ and $x$:
\begin{equation}\label{eq:p-from-eos-method} 
V_\mathrm{c}\int_{p_0}^{p}\frac{\mathrm{d}p'}{\tilde\rho(p')} =\int_{x_0}^x\mathrm{d}x'\,f_\mathrm{u}(x')=u(x_0)-u(x)\,,
\end{equation}
where $p_0=p(x_0)$ is determined by one of the conditions stated in Sec.~\ref{sec:meth-coar}.
In the latter case we proceed analogously via separation of variables $\rho$ and $x$.

The EOS method is particularly useful if long-range interactions are present.
For such cases an exact DFT calculation of density profiles tends to be be impracticable and calculations based on recursion relation for partition functions \cite{Davis:1990, Buschle/etal:2000a} are cumbersome. 
For long-range first-neighbor interactions, the SIVP method \cite{Bakhti/etal:2014} provides a user-friendly way to derive the EOS.

The microstates are encoded in a sequence of $N-1$ vacancies of size $m$ $(m=0,1,2,\ldots)$ between consecutive rods, $N$ in number. 
Summing over all microstates means summing over all size combinations of $N-1$ vacancies. 
This sum is free of constraints. 
The interaction energy of first-neighbor rods at distance $m$ is equivalent to part of the excitation energy $\epsilon_m(p)$ of the vacancies,
\begin{equation}\label{eq:epsilon} 
 \epsilon_m(p)=p m+\phi_m,
\end{equation}
where $\phi_m=\mathcal{V}_{i,i+m+\sigma}$. 
The vacancies themselves are free of interaction energies and form a set of polydisperse quasiparticles with generalized exclusion statistics. 
Their statistical mechanics has been worked out in \cite{Bakhti/etal:2014} building on a host of foundational work \cite{Haldane:1991, Wu:1994, Isakov:1994, Anghel:2007, Liu/etal:2011, Liu/etal:2012}. 
This treatment produces exact results for any thermodynamic quantity of interest for spatially homogeneous situations. 

The free enthalpy $G(p)$ per site is given by
\begin{equation}\label{eq:14} 
\beta G(p)=-\ln\left(1+\sum_{m=1}^\infty e^{-\beta\epsilon_m(p)}\right),
\end{equation}
from which other thermodynamic quantities are inferred via the auxiliary quantities \cite{Bakhti/etal:2014} 
\begin{equation}\label{eq:bij} 
 B_{\alpha\gamma}(p)\doteq\sum_{m=0}^\infty m^\alpha
 [\beta\epsilon_m(p)]^\gamma e^{-\beta\epsilon_m(p)}.
\end{equation}
This includes the mean size $\bar{m}$ of vacant cells 
\begin{align}
\label{eq:m-sivp} 
\bar{m}(p)&=\frac{B_{10}(p)}{B_{00}(p)}\,,
\end{align}
from which the EOS is obtained in the form
\begin{align}
\label{eq:eos-sivp} 
\tilde{\rho}(p)&=\frac{1}{\sigma+\bar{m}(p)}\,.
\end{align}

\subsection{Local pressure from pair density function}\label{sec:meth-loca}
In continuum mechanics, the `microscopic' pressure is defined as one third of the (negative) trace of the microscopic stress tensor and the divergence of this tensor determines the time evolution of the momentum density caused by the internal interaction forces. 
In one-dimension, the local stress tensor reduces to the microscopic pressure $p_\mathrm{mic}(x,t)$ and the
equation of motion for the momentum density $\Pi(x,t)$ becomes
\begin{align}\label{eq:momentum-flow}
\frac{\partial\Pi}{\partial t}&=
-\frac{\partial p_\mathrm{mic}}{\partial x}+f^\mathrm{ext}.
\end{align}
How statistical mechanical expressions for the local pressure $\bar{p}_\mathrm{mic}(x)=\langle p_\mathrm{mic}(x,t)\rangle_\mathrm{eq}$ in thermodynamic equilibrium (or the equilibrium-averaged microscopic stress tensor) are obtained when starting with these Euler equations of continuum mechanics was first studied by Irving and Kirkwood in 1950 \cite{Irving/Kirkwood:1950}. 
In Appendix~\ref{sec:appa} we have adapted the elegant derivation by Lutsko \cite{Lutsko:1988} to one dimension, which for pair interactions yields
\begin{align}\label{eq:p-rho2}
\bar{p}_\mathrm{mic}(x)=k_\mathrm{B}T\rho(x)+\!\!
\int_0^x\!\!\mathrm{d}x_1\!\!\int_x^L\!\!\mathrm{d}x_2\,\rho^{(2)}(x_1,x_2)
f(x_1,x_2)\,.
\end{align}
Here $\rho^{(2)}(x_1,x_2)$ is the pair distribution function and $f(x_1,x_2)$ the force of a particle at position $x_1$ on a particle at position $x_2$. The first term represents the kinematic pressure and the second the interaction pressure.
We set $V_\mathrm{c}=1$ throughout Sec.~\ref{sec:meth-loca}.

For hard rod systems, the interaction potential $v=v(|x_2-x_1|)$ is infinite for $|x_2-x_1|<\sigma$. 
This singularity can be accounted for in Eq.~(\ref{eq:p-rho2}) by considering a modified continuous potential $v_\epsilon(r)$, which agrees with $v(r)$ for $r\ge\sigma$ while for $r<\sigma$ it is given by 
$v_\epsilon(r)=v_0$ for $r\le\sigma-\epsilon$ and $v_\epsilon(r)=v(\sigma)-[(v_0-v(\sigma))/\epsilon](r-\sigma)$ for $\sigma-\epsilon\le r\le\sigma$.
After inserting the corresponding force in Eq.~(\ref{eq:p-rho2}) the pressure is obtained by taking the limit $v_0\to\infty$, $\epsilon\to0$ . For non-interacting hard rods ($v(r)=0$ for $r\ge\sigma$) in particular, this procedure yields
\begin{align}\label{eq:p-rho2-non-interacting}
\bar{p}_\mathrm{mic}(x)=k_\mathrm{B}T\rho(x)+
k_\mathrm{B}T\int_{x-\sigma}^{x}\!\!\mathrm{d}x'\,\rho^{(2)}(x',x'+\sigma)\,.
\end{align}
Because  $\rho^{(2)}(x,y)=0$ for $x<0$ or $y>L-\sigma$, the range of integration in the second term extends from zero to $x$ for $x\le\sigma$, and from $x-\sigma$ to $L-2\sigma$ for $(L-2\sigma)\le x\le(L-\sigma)$. 
This means that the interaction pressure at $x=0$ and $x=L-\sigma$ (the effective system boundary) vanishes. 
Accordingly, the local pressure at these boundary points is just given by the kinematic pressure, as earlier pointed out by Ibsen {\it et al.} \cite{Ibsen/etal:1997}, who derived Eqs.~(\ref{eq:p-rho2}) and (\ref{eq:p-rho2-non-interacting}) based on the approach in \cite{Irving/Kirkwood:1950}. 
For interacting hard rods ($V(r)\ne0$ for $r\ge\sigma$), the interaction pressure generally does not vanish at the system boundaries.

It is interesting to note that equating the expressions for the local pressure in Eqs.~(\ref{eq:pressure-profile-dft}) and (\ref{eq:p-rho2}) [or Eq.~(\ref{eq:p-rho2-non-interacting})] yields an integral equation connecting the pair distribution with the density. 
This could in principle be used to determine $\rho^{(2)}(x,y)$. 
Alternatively, the pair distribution can be obtained by solving the inhomogeneous Ornstein-Zernike relation with the direct correlation function given by the second order derivatives of the density functional. 
For hard rods with first-neighbor interactions, many-particle-density distribution functions are most conveniently obtained by employing recursion relations for partition functions, both for continuum \cite{Davis:1990} and for lattice systems \cite{Buschle/etal:2000a}. 
We note that for the latter, the integrals for the interaction pressure in Eqs.~(\ref{eq:p-rho2}) and (\ref{eq:p-rho2-non-interacting}) can be replaced by corresponding sums (Appendix \ref{sec:appc}).

%
\section{Gravitational field}\label{sec:grav}
%
Consider a semi-infinite vertical column of cells numbered $i=1,2,\ldots$ from the bottom up.
A uniform gravitational field $g$ acts on rods of mass $m_\mathrm{r}$.
We use it here to represent any linear potential.
We begin with non-interacting rods of size $\sigma$ on a lattice and then take the continuum limit.
Results from SIVP operating on a mesoscopic length scale are compared with those from DFT operating on a microscopic length scale.
Pressure profiles on a microscopic length scale obtained from DFT via (\ref{eq:pressure-dft-direct2}) are then compared with average microscopic pressure profiles inferred via (\ref{eq:p-rho2-non-interacting}).
Finally, we discuss some effects of repulsive contact interaction as made manifest in one or the other method.
For the sake of brevity we limit the discussion to one simple case study of each approach.
They can both be adapted to different applications.

\subsection{SIVP approach}\label{sec:grav-sivp}
The gravitational potential to be used in (\ref{eq:pressure-density-relation}) is
\begin{equation}\label{eq:10}
u(z)= m_\mathrm{r}gz.
\end{equation}
Convenient scaled variables for position, pressure, and temperature in this application are
\begin{equation}\label{eq:1} 
\hat{z}\doteq\frac{z}{z_\mathrm{s}}, \quad \hat{p}\doteq\frac{p}{p_\mathrm{s}}, \quad 
\hat{T}\doteq\frac{k_BT}{p_\mathrm{s}\sigma V_\mathrm{c}},
\end{equation}
where $z_\mathrm{s}=N\sigma V_\mathrm{c}$ is the length of all $N$ rods stacked up in a solid column.
The pressure at $z=0$ is $p_\mathrm{s}= Nm_\mathrm{r}g$, independent of $T$. 
The thermal energy $k_BT$ is measured in units of the work $p_\mathrm{s}\sigma V_\mathrm{c}$ required to lift this weight a distance equal to the size of one rod.

The SIVP analysis (Sec.~\ref{sec:meth-eosm}) starts from the expression for the density of vacant cells,
\begin{equation}\label{eq:21} 
 \bar{m}=\left[\exp\left(\frac{\hat{p}}{\sigma\hat{T}}\right)-1\right]^{-1},
\end{equation}
derived in \cite{Bakhti/etal:2014} from (\ref{eq:m-sivp}).
The scaled mass density (volume fraction) inferred from (\ref{eq:eos-sivp}) then reads
\begin{equation}\label{eq:22} 
\rho^{(\mathrm{mes})}=\sigma\rho=\frac{\exp\left(\frac{\hat{p}}{\sigma\hat{T}}\right)-1}
 {\exp\left(\frac{\hat{p}}{\sigma\hat{T}}\right)-1+\frac{1}{\sigma}}.
\end{equation}
Performing the integral (\ref{eq:p-from-eos-method}) with (\ref{eq:10}) and (\ref{eq:22}) yields the following equation for the pressure profile 
\begin{align}\label{eq:23} 
 \exp\left(\frac{(\sigma-1)(\hat{p}-1)}{\sigma\hat{T}}\right)
 \frac{e^{\hat{p}/\sigma\hat{T}}-1}{e^{1/\sigma\hat{T}}-1}=e^{-\hat{z}/\hat{T}}. 
\end{align}
The density profile follows from (\ref{eq:22}) by substitution.

\begin{figure}[t]
  \begin{center}
 \includegraphics[width=43mm]{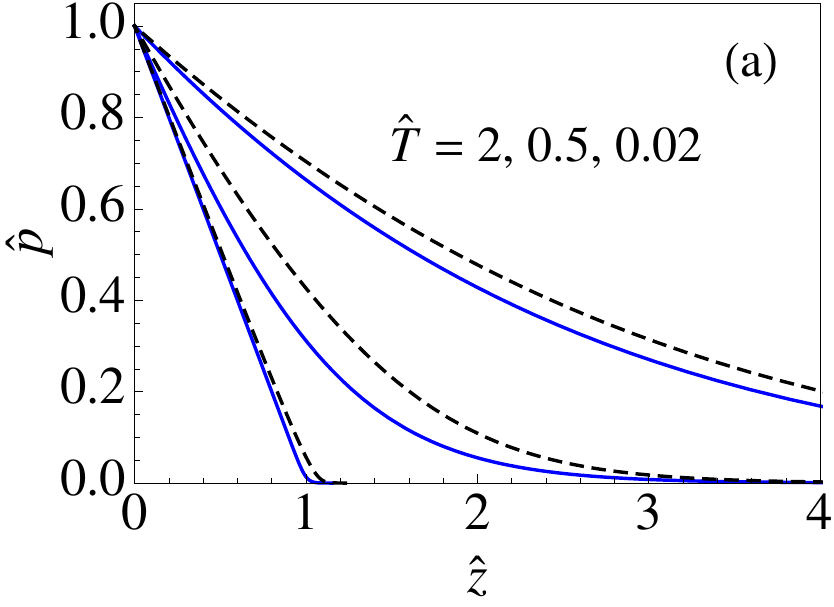}%
 \includegraphics[width=43mm]{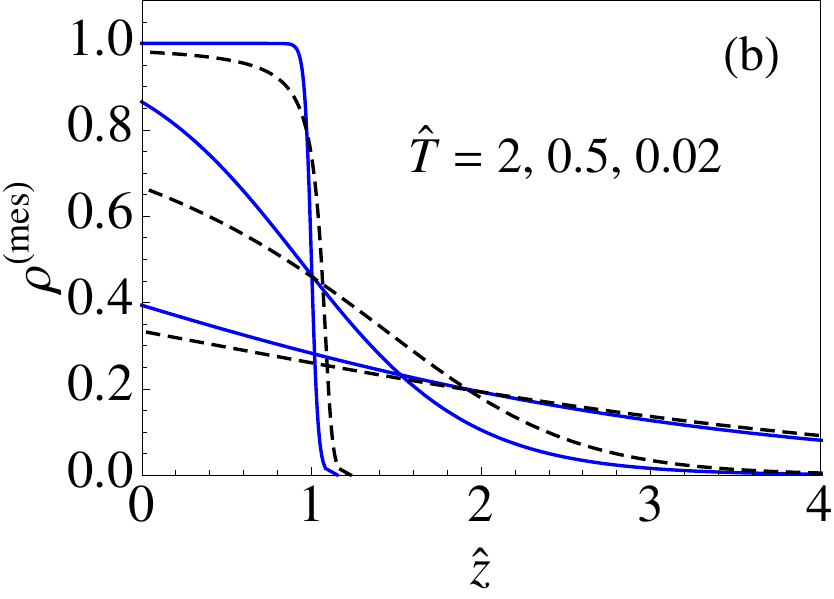}
\end{center}
\caption{(Color online) Profiles of (a) pressure and (b) density of rods in a uniform gravitational field at  different temperatures. 
The solid lines represent rods of size $\sigma=1$ on a lattice and the dashed lines rods (of any size) in a continuum.}
  \label{fig:fig1}
\end{figure}

The solid curves in Fig.~\ref{fig:fig1} show the profiles $\hat{p}(\hat{z})$ and $\rho^{(\mathrm{mes})}(\hat{z})$ at various values of $\hat{T}$ for rods of size $\sigma=1$.
The variation of pressure with height crosses over from hydrostatic to atmospheric with increasing temperature.
The density profile is rectangular in the low-temperature limit and varies like $\rho^{(\mathrm{mes})}\leadsto\hat{p}/\hat{T}$ at $\hat{T}\gg1$.
These profiles do not vary much with the rod size $\sigma$ when expressed by the scaled quantities (\ref{eq:1}).
In the limit $\sigma\to\infty$, $V_\mathrm{c}\to0$ with $\sigma V_\mathrm{c}=V_\mathrm{r}$ fixed, we have a system of rods of size $V_\mathrm{r}$ and mass $m_\mathrm{r}$ in a continuum.
The continuum versions of (\ref{eq:22}) and  (\ref{eq:23}) read
\begin{equation}\label{eq:25} 
\rho^{(\mathrm{mes})}=\frac{1}{\hat{T}/\hat{p}+1}, \quad \hat{p}-1+\hat{T}\ln\hat{p}=-\hat{z},
\end{equation}
respectively.
The pressure and mass density profiles in the continuum are shown as dashed curves in Fig.~\ref{fig:fig1}.

\subsection{DFT approach}\label{sec:grav-dft}
On the length scale of single rods additional features, not resolved by any EOS method including SIVP, emerge in the profiles for $\sigma\geq2$ when analyzed via DFT. 
We write
\begin{equation}\label{eq:102} 
\mathcal{U}_i=m_\mathrm{r}gz_i,\quad z_i=\left(i-\frac{1}{2}\right)V_\mathrm{c},
\end{equation}
where $z_i$ is the position of the center of the lowest cell occupied by a rod.
The free energy functional (\ref{eq:freeenergy-dft}) with no interaction except hardcore repulsion acquires the form (\ref{eq:b1}) and the density profile of rods for any given $\mathcal{U}_i$ are the solutions of the coupled equations (\ref{eq:b2}).
Here, for the linear potential (\ref{eq:102}), we set $M\to\infty$ in all expressions imported from Appendix~\ref{sec:appb}.

We first examine the case $\sigma=1$.
The solution (\ref{eq:b4}) is constructed from an exponential function as follows:
\begin{equation}\label{eq:19}
\bar{n}_i=\frac{\zeta\lambda_i}{1+\zeta\lambda_i},\quad \lambda_i=e^{-\hat{z}_i/\hat{T}},
\end{equation}
where $\hat{z}_i=z_i/\langle N\rangle\sigma V_\mathrm{c}$ and the average number $\langle N\rangle$ of rods is controlled by the fugacity $\zeta=e^{\hat{\mu}/\hat{T}}$, where $\hat{\mu}\doteq\mu/p_\mathrm{s}\sigma V_\mathrm{c}$ is the scaled chemical potential.
This DFT result exactly reproduces the functional dependence of $\rho^{(\mathrm{mes})}$ on $\hat{z}$ obtained via SIVP and given by (\ref{eq:22}) with (\ref{eq:23}) for $\sigma=1$ if we use (\ref{eq:101}) and set
\begin{equation}\label{eq:32} 
\zeta=e^{1/\hat{T}}-1.
\end{equation}
The results from both methods are fully consistent.
The SIVP solution remains exact even for small numbers of rods.
There exist no microscopic features in the density profile that SIVP does not resolve. 

Now we turn to the case $\sigma=2$, where the microscopic length scale does indeed reveal additional structures in the various profiles.
These structures are encoded in Eqs.~(\ref{eq:b5}), which for slowly varying profiles we expand into the form,
\begin{equation}\label{eq:108} 
\zeta\lambda_i=\frac{\bar{n}_i(1-\bar{n}_i)}{(1-2\bar{n}_i)^2}\left[1+\frac{\bar{n}_{i-1}-2\bar{n}_i+\bar{n}_{i+1}}{1-2\bar{n}_i}+\cdots\right],
\end{equation}
and note that the first correction is of second order.
The leading term alone leads to the density profile,
\begin{equation}\label{eq:34} 
\rho_i^{(\mathrm{mes})}\doteq 2\bar{n}_i=1-\frac{1}{\sqrt{1+4\zeta \lambda_i}},
\end{equation}
which coincides with (\ref{eq:22}) and (\ref{eq:23}) for $\sigma=2$ if we set
\begin{equation}\label{eq:35} 
\zeta=e^{1/2\hat{T}}\big(e^{1/2\hat{T}}-1\big).
\end{equation}
Thus full consistency between the two approaches is established on the mesoscopic length scale.

Finding the microscopic structures in the density profiles of rods and mass requires that we solve Eqs.~(\ref{eq:b5}) and then use (\ref{eq:101}) instead of Eq.~(\ref{eq:34}).
The solution as derived in Appendix~\ref{sec:appb} reads
\begin{equation}\label{eq:113} 
\bar{n}_i=\sum_{l=0}^\infty(-1)^l\prod_{k=0}^l\frac{h_{i+k}}{1+h_{i+k}},\quad i=1,2,\ldots,
\end{equation}
with the $h_i$ determined recursively from 
\begin{equation}\label{eq:111} 
h_0=0,\quad h_i=\frac{\zeta\lambda_i}{1+h_{i-1}},\quad i=1,2,\ldots.
\end{equation}

\begin{figure}[t]
  \begin{center}
 \includegraphics[width=43mm]{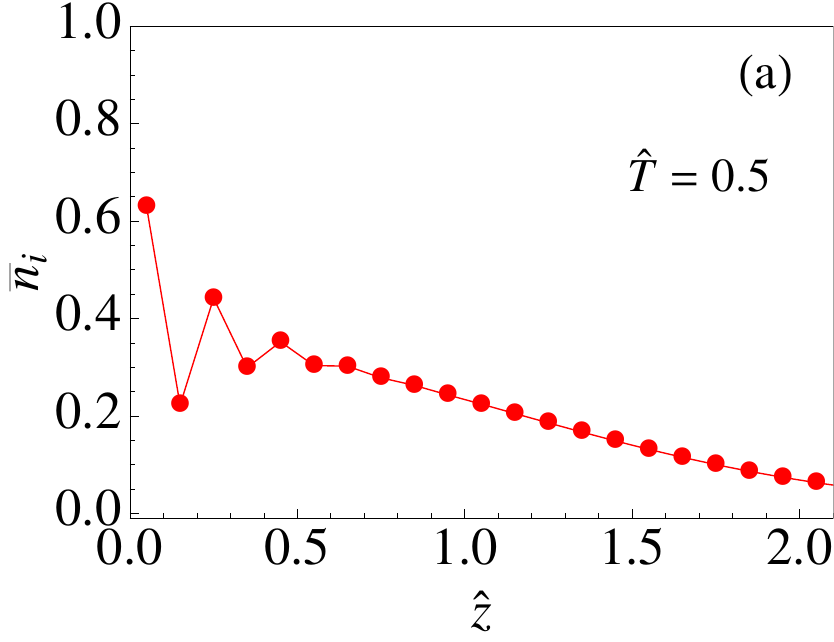}%
 \includegraphics[width=43mm]{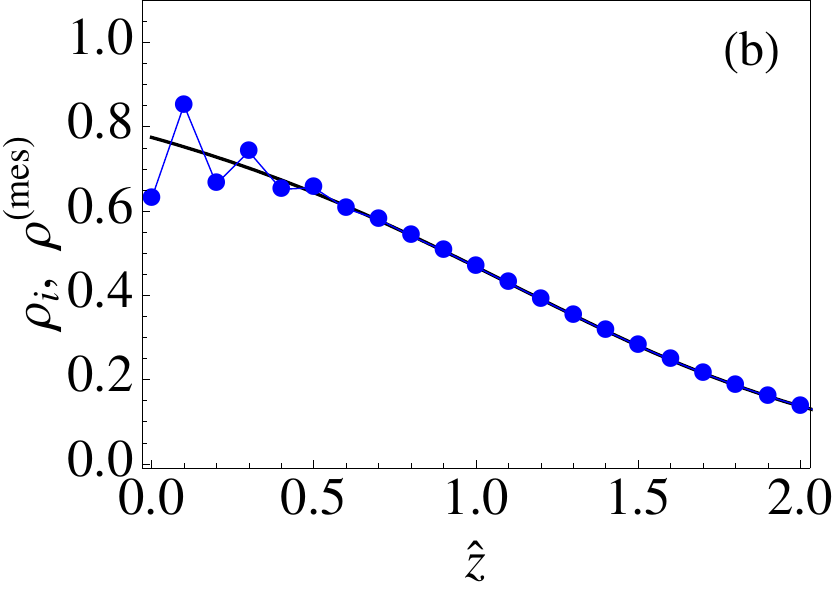}
 \includegraphics[width=43mm]{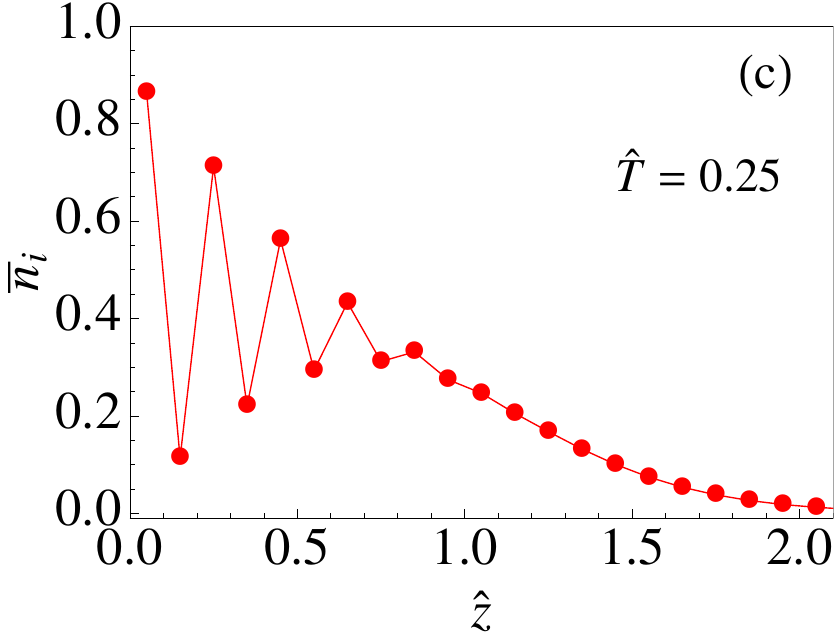}%
 \includegraphics[width=43mm]{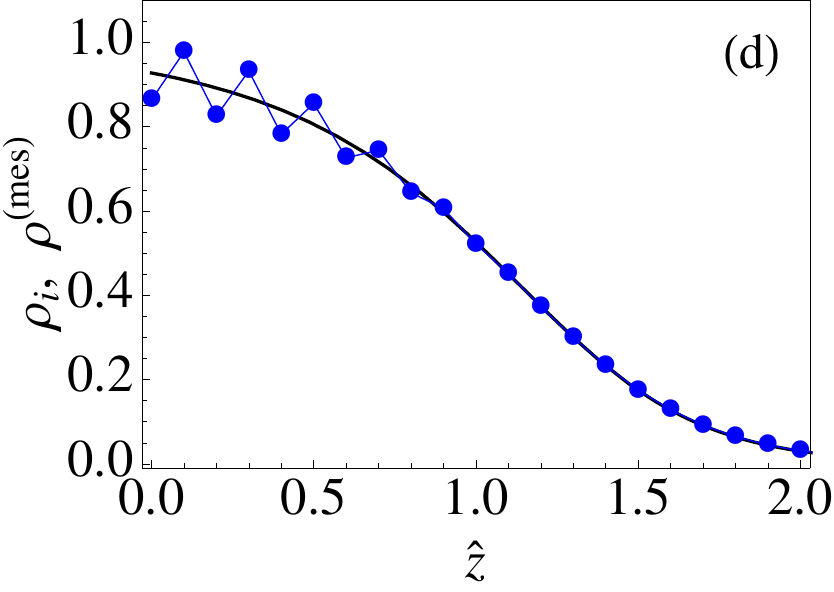}
 \includegraphics[width=43mm]{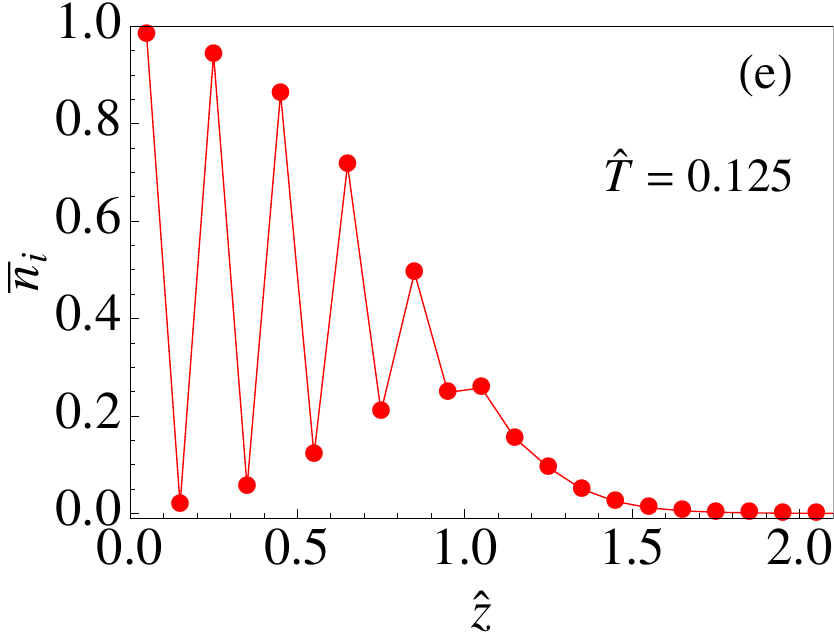}%
 \includegraphics[width=43mm]{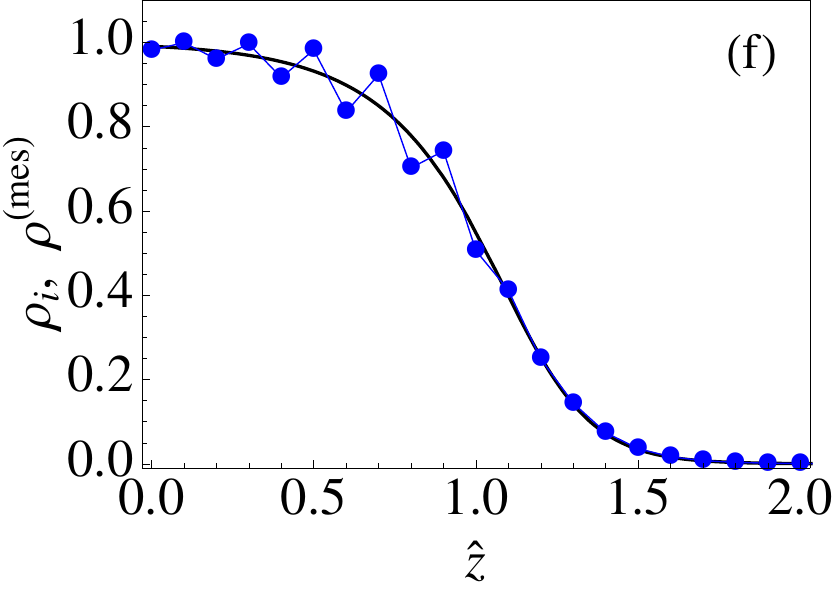}
  \includegraphics[width=43mm]{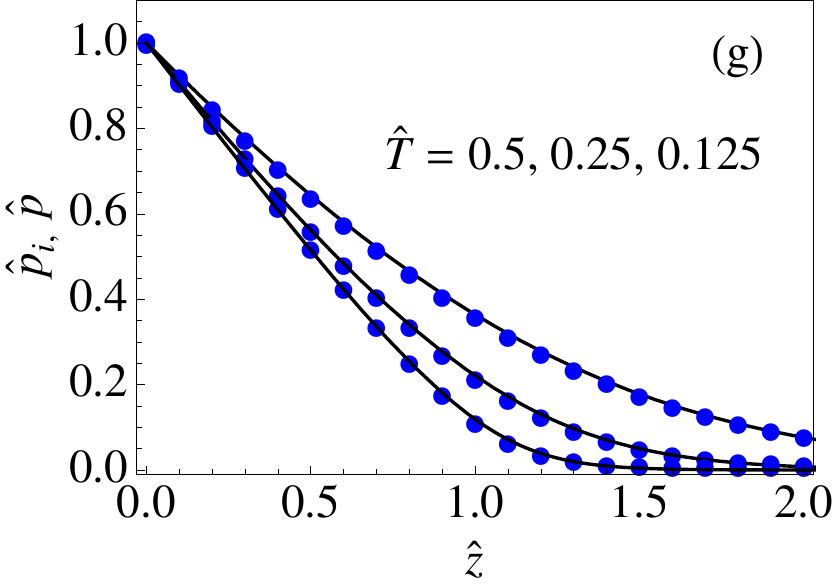}
\end{center}
\caption{(Color online) Density and pressure profiles for rods of size ${\sigma=2}$ in a uniform gravitational field at  different temperatures. 
The solid curves represent solutions of (\ref{eq:22}) and (\ref{eq:23}) as predicted by SIVP for a system with $N\gg1$.
The $\bar{n}_i$-data on the left (circles) originate from (\ref{eq:113}) with $\zeta$ from (\ref{eq:35}) and $\lambda_i$ from (\ref{eq:19}) with $\langle N\rangle=5$. 
These data are transcribed into the circles on the right via (\ref{eq:101}) and into the circles of panel (g) via (\ref{eq:305}).}
  \label{fig:fig9}
\end{figure}

The circles in the panels on the left in Fig.~\ref{fig:fig9} are derived from (\ref{eq:113}).
The probability $\bar{n}_i$ that a rod occupies cells $i$ and $i+1$ varies with index $i$ in a manner that reflects the combined effects of the hardcore exclusion interaction between rods and the presence of a hard floor $\hat{z}=0$.
The spatial oscillations are mild at high $T$ and discernible only very close to the floor.
As $T$ is lowered, the amplitude becomes stronger and the range wider. 
In the limit $T\to0$ the $\bar{n}_i$ strictly alternate between one and zero, reflecting a compact stack of rods.

When transcribed via (\ref{eq:101}) to the mass density, the wall effect is not nearly as strong.
The evidence is represented by the circles in the panels on the right, where we set $\hat{z}=\frac{1}{2}(\hat{z}_i+\hat{z}_{i-1})$ for the discrete data and $\hat{z}$ from (\ref{eq:1}) for the curves.
At high $T$ the effect is still strongest in the immediate vicinity of the floor but that is no longer the case at the lowest $T$ used in Fig.~\ref{fig:fig9}. 
At $T=0$ the effect disappears altogether. 

The local pressure predicted by DFT as inferred from (\ref{eq:pressure-dft-direct2}) with the $\bar{n}_i$ from (\ref{eq:b2}) substituted into the free-energy functional (\ref{eq:b1}) becomes
\begin{equation}\label{eq:305} 
p_iV_\mathrm{c}=k_\mathrm{B}T\ln\left(1+\frac{\bar{n}_i}{1-\rho_i}\right)
\end{equation}
with $\rho_i$ from (\ref{eq:101}).

In panel (g) of Fig.~\ref{fig:fig9} we show the pressure profiles thus obtained from the density data of panels (a)-(f). 
The profiles are monotonically decreasing  from $\hat{p}(0)=1$. 
Remarkably, the wild oscillations of the $\bar{n}_i$ are almost completely smoothed by (\ref{eq:305}). 
However, there does exist a systematic albeit small deviation between the DFT and SIVP pressure profiles. 
The DFT pressure profile is closer to but not identical with the true  microscopic pressure as will be further discussed in Sec.~\ref{sec:grav-micro}.

The agreement of the DFT and SIVP approaches on the mesoscopic length scale defined earlier is underlined by expressions (\ref{eq:19}) and (\ref{eq:32}) for $\sigma=1$ and by expressions (\ref{eq:34}) and (\ref{eq:35}) for $\sigma=2$.
It can further be shown that for any $\sigma$ the density $\rho^{(\mathrm{mes})}(\hat{z})$ determined by (\ref{eq:22}) and (\ref{eq:23}) is the solution of a polynomial equation of order $\sigma$.
Likewise, the probabilities $\bar{n}_i$ at $\hat{z}_i$ inferred from (\ref{eq:b2}) with all $\bar{n}_j$ within the square brackets set equal to each other is also the solution of a polynomial equation of order $\sigma$.
The control variables are $\hat{T}$, $N$ in the first polynomial equation (canonical ensemble) whereas they are $\hat{T}$ and $\zeta$ in the second polynomial equation (grandcanonical ensemble).
The two polynomial equations are equivalent if we set 
\begin{equation}\label{eq:89} 
\rho_i^{(\mathrm{mes})}=\sigma \bar{n}_i,\quad \zeta=e^{(\sigma-1)/\sigma\hat{T}}\big(e^{1/\sigma\hat{T}}-1\big).
\end{equation}
The agreement between DFT and SIVP on the mesoscopic length scale also extends to the pressure profiles.
From (\ref{eq:305}) with $\rho_i^{(\mathrm{mes})}=\sigma\bar{n}_i$ we infer
\begin{equation}\label{eq:306} 
\frac{\hat{p}}{\sigma\hat{T}}=\ln\left(1+\frac{\rho^{(\mathrm{mes})}}{\sigma(1-\rho^{(\mathrm{mes})})}\right),
\end{equation}
which is equivalent to (\ref{eq:22}).

The microscopic features in the density profile remain conspicuous for $\sigma>2$. 
The oscillations that are superimposed on profile predicted by SIVP are characterized by a `wavelength' proportional to $\sigma$.
Such profiles are readily produced from (\ref{eq:b15}) and (\ref{eq:b16}).
The exact hard-wall effects for rods of arbitrary size $\sigma$ will be investigated in Sec.~\ref{sec:ster}. 
Soft walls as realized in power-law traps also produce structures on a microscopic length scale. 
Some examples will be investigated in Sec.~\ref{sec:powe}. 

\subsection{Average local microscopic pressure}\label{sec:grav-micro}
\begin{figure}[t]
  \begin{center}
 \includegraphics[width=70mm]{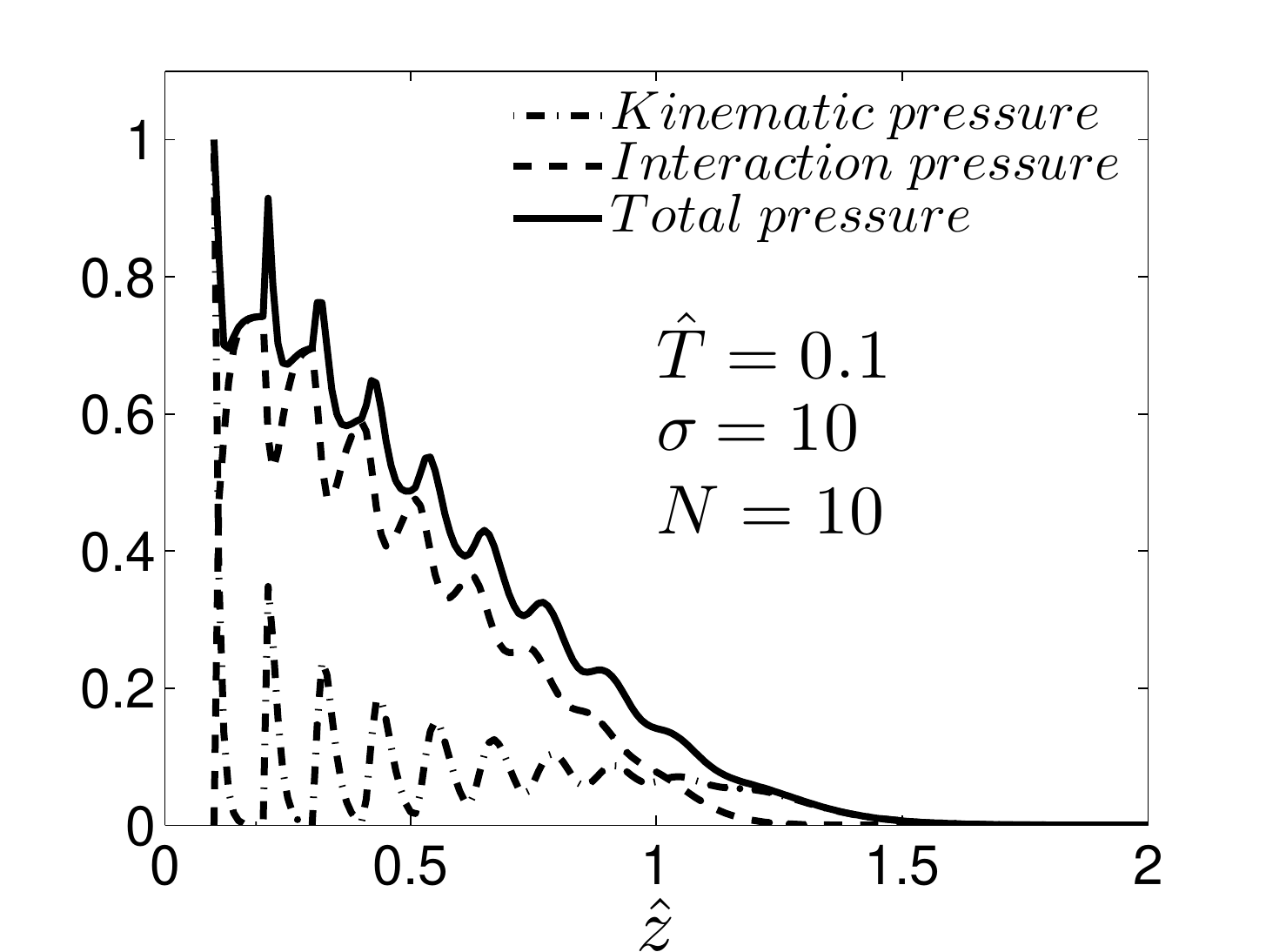}
\end{center}
\caption{Profiles of kinematic, interaction and total pressure $\bar{p}_\mathrm{mic}$ for rods of size $\sigma=10$ at low $\hat{T}$ as inferred from (\ref{eq:p-rho2-non-interacting}) and the pair distribution function as derived in Appendix \ref{sec:appc}.}
  \label{fig:microscopic_pres}
\end{figure}

\begin{figure}[t]
  \begin{center}
 \includegraphics[width=42mm]{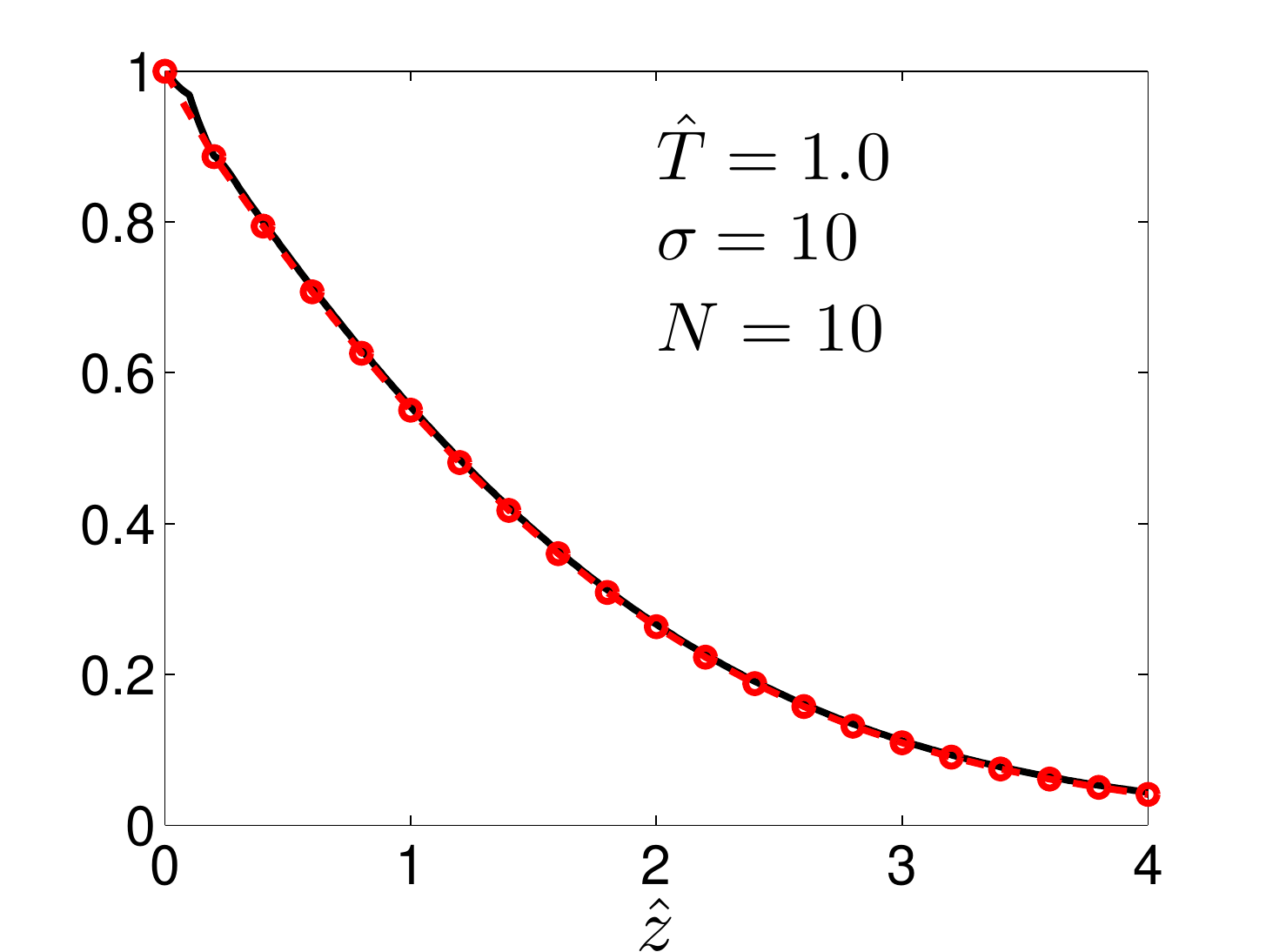}
 \includegraphics[width=42mm]{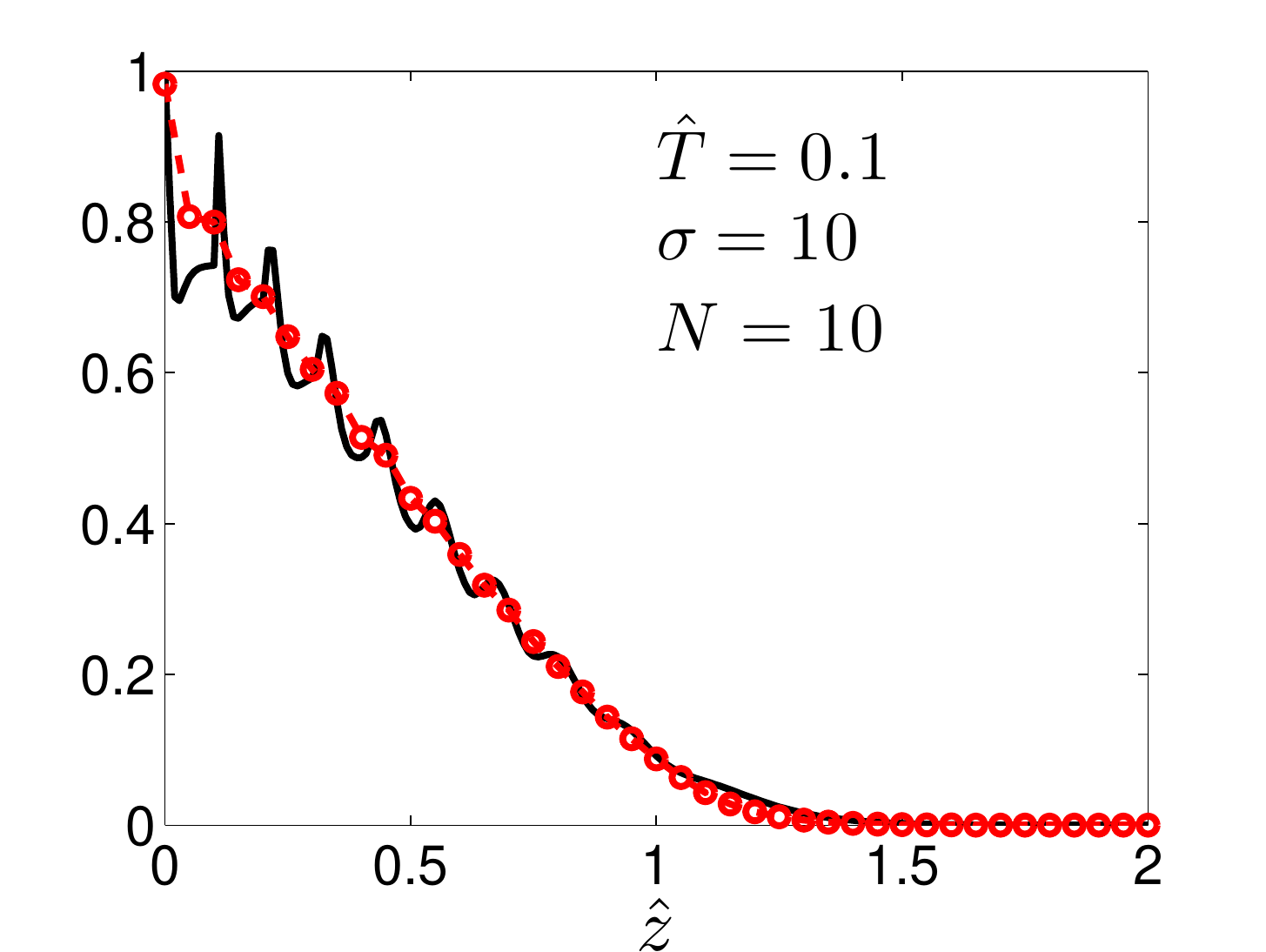}
\end{center}
\caption{(Color online) Profiles of the local pressure calculated from (\ref{eq:p-rho2-non-interacting}) in conjunction with (\ref{eq:micscopic_pressure}) (circles connected by dashed lines) and calculated from (\ref{eq:pressure-dft-direct2}) (solid lines)  for high $\hat{T}$ (left) and low $\hat{T}$ (right).}
  \label{fig:microscpic_dft_pres}
\end{figure}

The average local microscopic pressure $\bar{p}_\mathrm{mic}(x)$ as inferred via (\ref{eq:p-rho2-non-interacting}) from the pair distribution function by the method presented in Appendix~\ref{sec:appc} has two parts: a kinematic pressure and an interaction pressure (in a formal sense).
In Fig.~\ref{fig:microscopic_pres} we show the profiles of both parts and their sum  for rods of size $\sigma=10$ at low temperature $(\hat{T}=0.1)$.
Both parts show strong oscillations that are somewhat out of phase.
These oscillations are strongly attenuated with distance from the hard floor.
They quickly become imperceptible with rising $T$ (see Fig.~\ref{fig:microscpic_dft_pres}).
The non-monotonic features of $\bar{p}_\mathrm{mic}(x)$ remain totally unresolved in the EOS pressure profiles discussed earlier but are partially resolved by the DFT pressure profiles as illustrated in Fig.~\ref{fig:microscpic_dft_pres}.
At high $\hat{T}$ the two profiles are virtually identical and equal to the SIVP profile.
At low $\hat{T}$, the additional $\bar{p}_\mathrm{mic}(x)$ oscillations are reproduced by DFT with remarkable accuracy albeit not exactly \footnote{The microscopic pressure at $\hat{z}=0$ is not equal the total weight of the rods but it has the form $p_0 = N_rm_rg + p_L$, as can be shown from (\ref{eq:part_fct_explicit}), where 
$p_L$ is the pressure at the top of the system. With increasing system size $p_L$ approaches zero.}. 

\subsection{Contact interaction via SIVP}\label{sec:grav-sivp-cont}
\begin{figure}[t]
  \begin{center}
 \includegraphics[width=43mm]{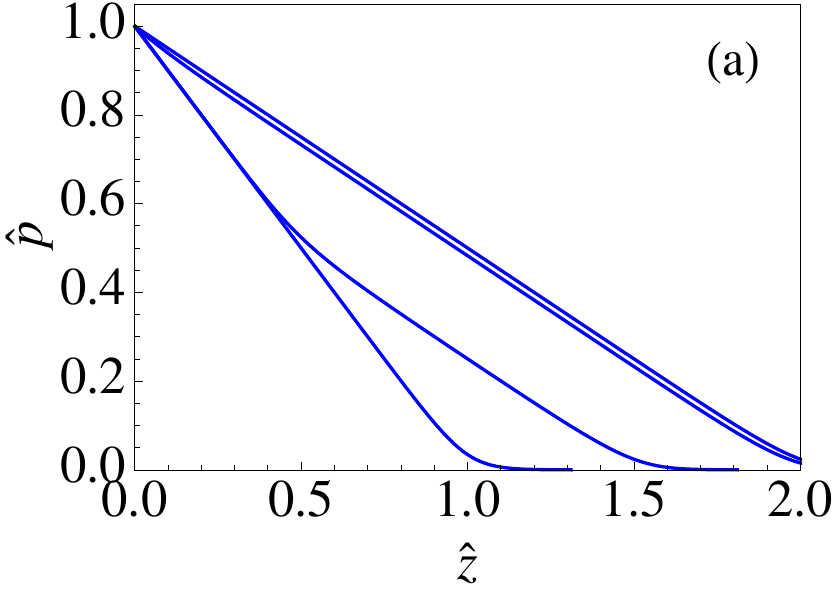}%
 \includegraphics[width=43mm]{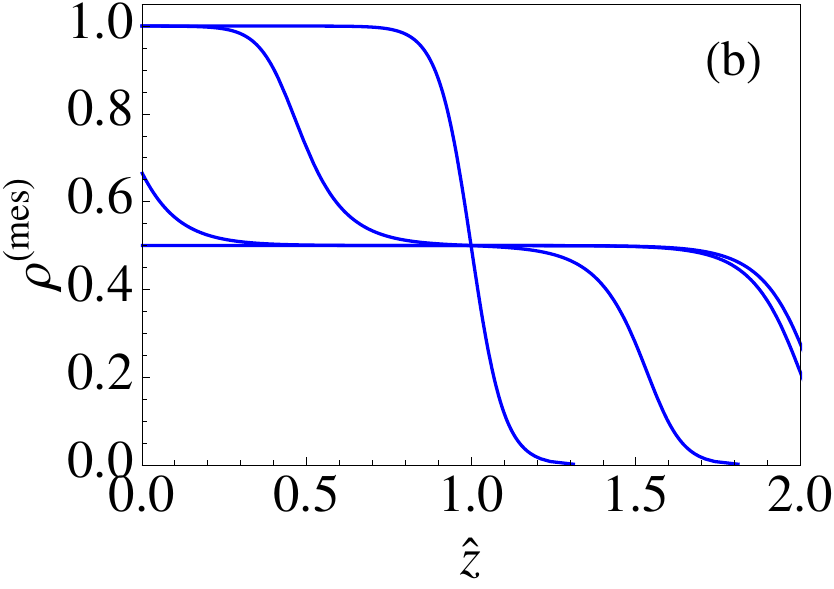}
 \includegraphics[width=43mm]{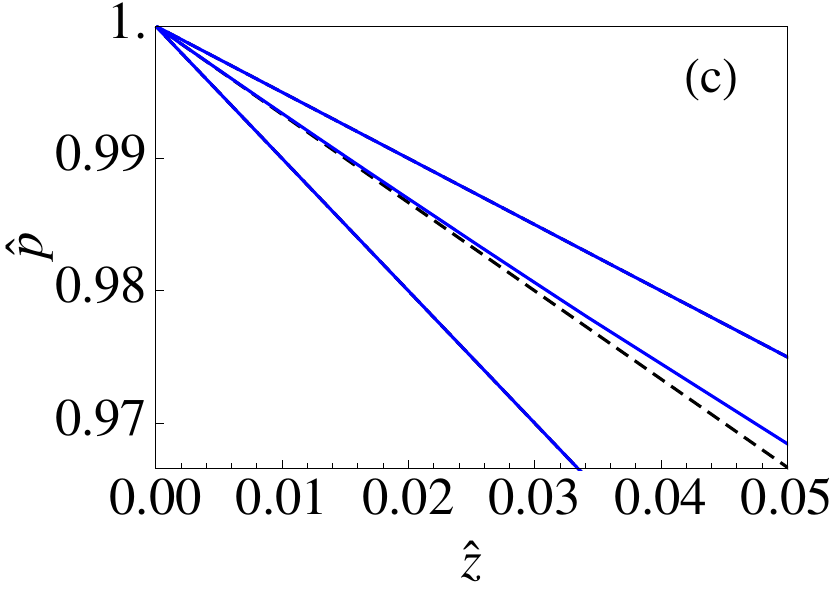}
\end{center}
\caption{(Color online) Profiles of (a) pressure and (b) density of rods (of size $\sigma=1$) in a uniform gravitational field at temperatures $\hat{T}=0.05$ and for (repulsive) contact interaction $\hat{v}=0, -0.5, -1, -5$. Panel (c) shows the $\hat{p}$ versus $\hat{z}$ data near $\hat{z}=0$. The dashed line has slope $-2/3$.}
  \label{fig:fig5}
\end{figure}

Consider a contact interaction potential $v$ that is attractive for $v>0$ and repulsive for $v<0$.
As inferred from \cite{Bakhti/etal:2014}, the contact interaction changes the local density of vacant cells from (\ref{eq:21}) to
\begin{equation}\label{eq:36} 
  \bar{m} = \frac{1}{(1-t^{\hat{p}})\big[1+(1-t^{\hat{p}})t^{-\hat{p}-\hat{v}}\big]},
\end{equation}
where we have introduced the variable
\begin{align}\label{eq:158} 
  t \doteq  e^{-1/\sigma\hat T}  \qquad (0\leq t \leq 1)
\end{align}
and the scaled interaction $\hat{v}\doteq v/\sigma p_\mathrm{s}V_\mathrm{c}$ in addition to the scaled variables (\ref{eq:1}).
The mass density $\rho^{(\mathrm{mes})}$ is inferred from (\ref{eq:eos-sivp}).
The integral (\ref{eq:p-from-eos-method}) can still be evaluated exactly and yields the pressure profile,
\begin{equation}\label{eq:38} 
 t^{\sigma(1-\hat{p})}
  \frac{1-t(1-t^{\hat{v}})}{1-t^{\hat{p}}(1-t^{\hat{v}})}
   \frac{1-t^{\hat{p}}}{1-t}
  =  t^{\sigma\hat{z}},
\end{equation}
in generalization of (\ref{eq:23}).
An attractive contact interaction affects the profiles in a way similar to what a drop in temperature does (see Fig.~\ref{fig:fig1}). 
No significant additional features make their appearance.
Repulsion is more interesting in that respect as documented in Fig.~\ref{fig:fig5} for $\sigma=1$.

The configuration of rods becomes stratified at low temperature.
The density profile now exhibits an additional layer of intermediate density.
The width of that layer increases with the strength of the repulsion.
The (largely hydrostatic) pressure profile acquires different slopes inside different layers.

These profiles depend systematically on $\sigma$ without producing any additional features. 
From \cite{Bakhti/etal:2014} we know that the effects of attractive or repulsive contact interactions of finite strength fade away completely in the continuum limit.

One subtle feature of note concerns the initial slope of the hydrostatic pressure as presented in panel (a) of Fig.~\ref{fig:fig1} and then again in panel (c) on a much expanded scale.
What in panel (a) looks like a clean change between slope $-1$ for interaction strengths $\hat{v}>-1$ and slope $-1/2$ for stronger contact repulsion, $\hat{v}\leq-1$ is, at very small $\hat{z}$, a two-step change with slope $-2/3$ at $\hat{v}=-1$ in the middle.
This asymptotic slope is evident in the expansion of (\ref{eq:38}):
\begin{align}\label{eq:159} 
  \hat{p} = 1 - \hat{z}\left[1-\frac{1}{2-t+t^{-\hat{v}-1}(1-t)^{2}}\right]+\mathrm{O}(\hat{z}^{2}).
\end{align}

\subsection{Contact interaction via DFT}\label{sec:grav-dft-cont}
Here we wish to demonstrate how a repulsive contact force of infinite strength effectively increases the size of rods.
The inclusion of a scaled contact potential ${v_\mathrm{c}=-\hat{v}/\hat{T}}$ generalizes (\ref{eq:b1}) to 
\begin{align}\label{eq:43}
\beta F&=\sum_{i=1}^L\left\lbrace
\left(1-\sum_{j=i-\sigma}^i\tilde{n}_j\right)\ln\left(1-\sum_{j=i-\sigma}^i\tilde{n}_j + C_{i-\sigma,i}\right)\right.\nonumber\\
&p_i\ln\left(\tilde{n}_i-C_{i-\sigma,i}\right)+\tilde{n}_{i-\sigma}\ln\left(\tilde{n}_{i-\sigma}-C_{i-\sigma,i}\right)\nonumber\\
& \hspace{-2mm}\left.- \bar{n}_{i-\sigma}\ln \tilde{n}_{i-\sigma}-\left(1-\sum_{j=i-\sigma}^{i-1}\tilde{n}_j\right)
\ln\left(1 -\sum_{j=i-\sigma}^{i-1}\tilde{n}_j\right)\right\rbrace
\end{align}
as shown in \cite{Bakhti/etal:2013}, where the contact interaction $v_\mathrm{c}$ is contained in the correlators,
\begin{equation}\label{eq:44}
C_{i-\sigma,i}= \frac{A_i-\sqrt{A_i^2-4e^{-v_c}(e^{-v_c}-1)\tilde{n}_{i-\sigma}\tilde{n}_i}}{2(e^{-v_c}-1)},
\end{equation}
\begin{equation}\label{eq:45}
A_i=1+e^{-v_c}(\tilde{n}_{i-\sigma}+\tilde{n}_i)-\sum_{k=i-\sigma}^{i}\tilde{n}_k,
\end{equation}
that appear in the first three terms of (\ref{eq:43}).
The last two terms reflect the hardcore repulsion.

The metamorphosis of hard rods of one size into hard rods of a bigger size is most  transparent if we consider the case $\sigma=1$ and compare the limits $v_\mathrm{c}=0$ and $v_\mathrm{c}=+\infty$.
The extremum condition (\ref{eq:structure-equations}) applied to (\ref{eq:43}) leads to the following set of relations
that determine the density profile $\{\bar{n}_i\}$ for external potential $\mathcal{U}_i$ and fugacity $\zeta$:
\begin{align}\label{eq:46}
&\zeta e^{-\beta \mathcal{U}_i} =\\
&\frac{[1-\bar{n}_i][\bar{n}_i-C_{i-1,i}][\bar{n}_i-C_{i,i+1}]}{\bar{n}_i[1-\bar{n}_{i-1}-\bar{n}_{i} + C_{i-1,i}][1-\bar{n}_i-\bar{n}_{i+1} + C_{i,i+1}]},\nonumber
\end{align}
\begin{equation}\label{eq:47}
C_{i-1,i}= \frac{A_i-\sqrt{A_i^2-4\eta(\eta +1)\bar{n}_{i-1}\bar{n}_i}}{2\eta},
\end{equation}
where $A_i=1+ \eta(\bar{n}_{i-1} +\bar{n}_i)$ and $\eta = e^{-v_c}-1$.
For $v_\mathrm{c}\to0$ we have $C_{i,i+1}=\bar{n}_i\bar{n}_{i+1}$ and (\ref{eq:46}) reduces to
\begin{align}\label{eq:48} 
\zeta e^{-\beta\mathcal{U}_i} =\frac{\bar{n}_i}{1-\bar{n}_i},
\end{align}
whereas for $v_\mathrm{c}\to\infty$ we have $C_{i,i+1}=0$ and (\ref{eq:46}) becomes
\begin{align}\label{eq:49} 
\zeta e^{-\beta\mathcal{U}_i} = \frac{\bar{n}_i(1-\bar{n}_{i})}{(1-\bar{n}_{i-1}-\bar{n}_{i})(1-\bar{n}_i-\bar{n}_{i+1})},
\end{align}
representing rods of size $\sigma=2$ with only hardcore repulsion.
In the application to a uniform gravitational field, (\ref{eq:48}) is equivalent to (\ref{eq:19}) and (\ref{eq:49}) is equivalent to (\ref{eq:b5}).

The SIVP approach of Sec.~\ref{sec:grav-sivp-cont} describes the same crossover from rods of size $\sigma$ to rods of size $\sigma+1$ under a repulsive contact interaction of increasing strength.
The pressure profile (\ref{eq:38}) evaluated for $\hat{v}=0$ and any $\sigma$ reproduces (\ref{eq:23}). 
When the same expression is evaluated for $\hat{v}=-\infty$ it connects again with (\ref{eq:23}) but now for $\sigma+1$ provided the scaled variables are properly adjusted.

%
\section{Power-law trap}\label{sec:powe}
%
Optical or magnetic traps of several different designs for atomic or molecular gases produce wells with a range of profiles.
How does the pressure at the center of the trap vary with temperature?
How does the shape of the trap potential affect the profiles of density and pressure?
Power-law traps are well-suited for our two approaches and can illuminate these questions with answers from an exact analysis.

\subsection{Profiles for lattice and continuum}\label{sec:powe-latt-cont}
In this application we consider an infinite row of cells numbered $i=0,\pm1,\pm2,\ldots$ at positions $x_i=iV_\mathrm{c}$.
The rods are confined to a region centered at $x=0$ by the symmetric power-law potential
\begin{equation}\label{eq:26} 
 u(x)=u_0\left|\frac{x}{x_0}\right|^\alpha, \quad \alpha>0
\end{equation}
with $u_0$ representing a depth and $x_0$ representing (at least for $\alpha>1$) a width of the trap.

The analysis proceeds as in Sec.~\ref{sec:grav-sivp}. 
Expression (\ref{eq:22}) remains unchanged. 
However, the pressure profile is now determined by the relation
\begin{align}\label{eq:27} 
 \exp\left(\frac{(\sigma-1)(\hat{p}-\hat{p}_T)}{\sigma\hat{T}}\right)
 \frac{e^{\hat{p}/\sigma\hat{T}}-1}{e^{\hat{p}_T/\sigma\hat{T}}-1}=e^{-|\hat{x}|^\alpha/\hat{T}}, 
\end{align}
where, in addition to $\hat{p}$ and $\hat{T}$ from (\ref{eq:1}), we use the scaled variables
\begin{equation}\label{eq:28}
\hat{x}\doteq\frac{x}{x_\mathrm{s}},\qquad \hat{p}_T\doteq\frac{p_T}{p_\mathrm{s}}.
\end{equation}
A solid stack of rods extends out to $x_\mathrm{s}= \frac{1}{2}N\sigma V_\mathrm{c}$ and the pressure at the center becomes $p_\mathrm{s}=u_0(x_\mathrm{s}/x_0)^\alpha/(\sigma V_\mathrm{c})$. 
In the continuum limit, $\sigma\to\infty$, $V_\mathrm{c}\to0$ with $\sigma V_\mathrm{c}=V_\mathrm{r}$, Eqs.~(\ref{eq:22}) and (\ref{eq:27}) turn into
\begin{equation}\label{eq:302} 
\rho^{(\mathrm{mes})}=\frac{1}{1+\hat{T}/\hat{p}},\quad \hat{p}_T-\hat{p}-\hat{T}\ln\frac{\hat{p}_T}{\hat{p}}=|\hat{x}|^\alpha.
\end{equation}

One additional relation is needed to bring closure to (\ref{eq:22}) and (\ref{eq:27}), namely
\begin{equation}\label{eq:29} 
\int_0^\infty d\hat{x}\,\rho^{(\mathrm{mes})}(\hat{x})=1,
\end{equation}
reflecting mass conservation.
The pressure $\hat{p}_T$ at $\hat{x}=0$ can be determined from this relation.
For the two extreme rod sizes we thus obtain
\begin{equation}\label{eq:303} 
\Gamma(1/\alpha+1)f_{1/\alpha}\big(e^{\hat{p}_T/\hat{T}}-1\big)\hat{T}^{1/\alpha}=1\quad (\sigma=1),
\end{equation}
\begin{equation}\label{eq:304} 
\frac{\hat{p}_T^{1/\alpha}}{\alpha}\int_0^1 dk\left[1-k-\frac{\hat{T}}{\hat{p}_T}\ln k\right]^{\frac{1-\alpha}{\alpha}}\!\!\!=1\quad 
(\sigma=\infty),
\end{equation}
where $f_{n}(z)$ is the Fermi-Dirac function.

The rods remain confined at any finite temperature: we have $\hat{p}_T>0$ if $\hat{T}<\infty$ for any $\alpha>0$. 
The limit $\alpha\to\infty$, representing a trap of width $2x_0$ with rigid walls, is subtle.
Relation (\ref{eq:303}) reduces to
\begin{equation}\label{eq:51} 
\Gamma(1)\left(1-e^{-\hat{p}_T/\hat{T}}\right)\frac{x_0}{x_\mathrm{s}}=1.
\end{equation}
We then have $\hat{p}=\hat{p}_T$ inside the trap, with a uniform density, $\rho^{(\mathrm{mes})}=x_\mathrm{s}/x_0$. 
We thus recover the familiar EOS $pV_\mathrm{c}/k_\mathrm{B}T=-\ln(1-\rho^{(\mathrm{mes})})$ of the ideal lattice gas.

The limit $\hat{T}\to0$ yields $\hat{p}_T=1$ and the profiles for pressure and density are
\begin{equation}\label{eq:50} 
\hat{p}=\big(1-|\hat{x}|^\alpha\big)\theta(1-\hat{x}),\quad \rho^{(\mathrm{mes})}=\theta(1-\hat{x}),
\end{equation}
respectively.
If $\alpha>1$ $(\alpha<1)$ the pressure $\hat{p}_T$ at the center of the trap increases (decreases) with $\hat{T}$ rising from zero.
For the linear potential $(\alpha=1)$, which is equivalent to the case of the uniform gravitational field (Sec.~\ref{sec:grav-sivp}), we  have $\hat{p}_T=1$ for all $\hat{T}$.

\begin{figure}[b!]
  \begin{center}
 \includegraphics[width=43mm]{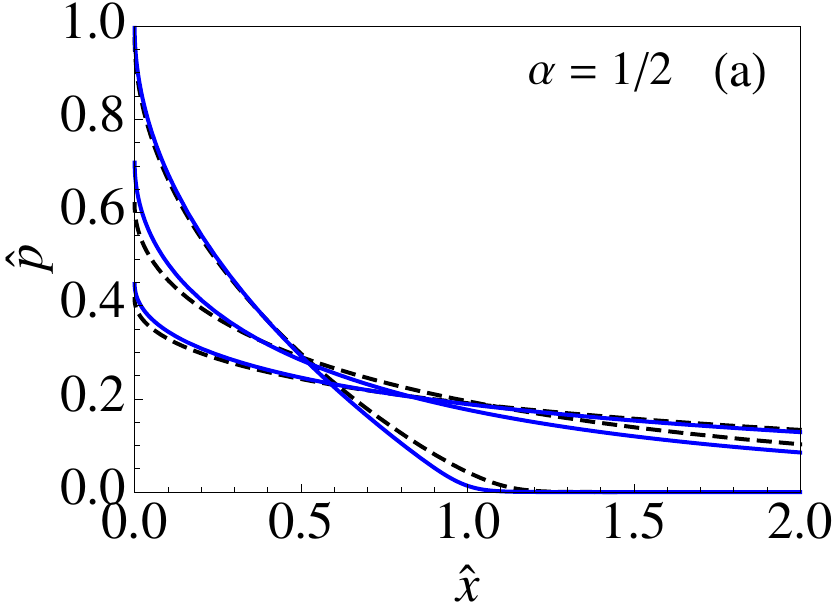}%
 \includegraphics[width=43mm]{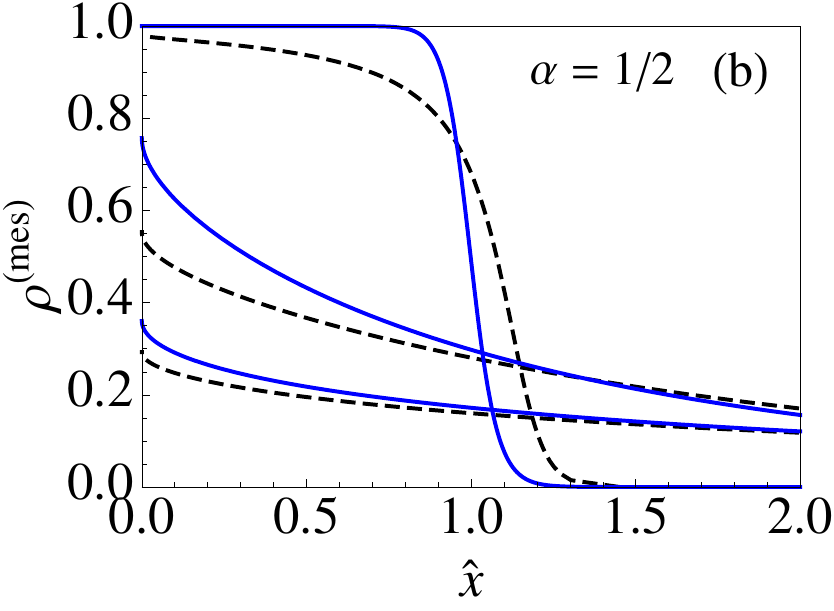}
  \includegraphics[width=43mm]{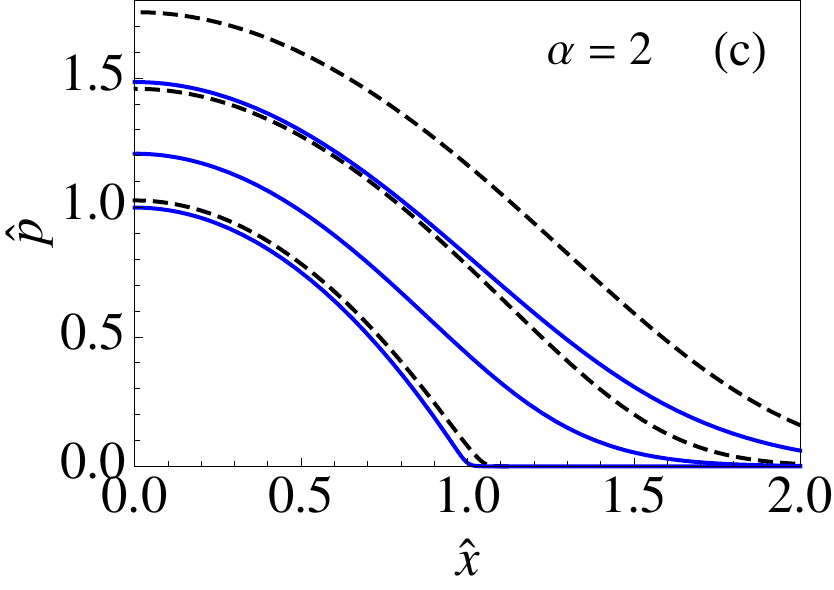}%
 \includegraphics[width=43mm]{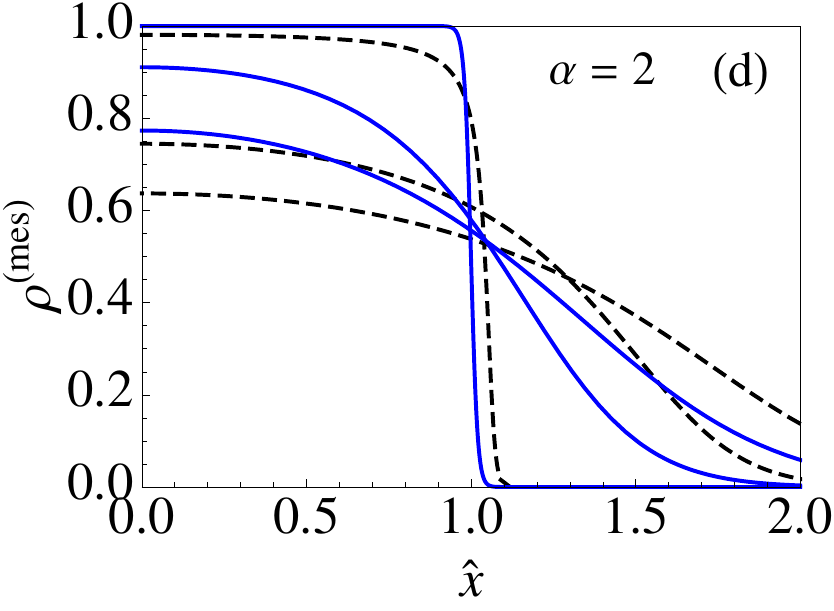}
\end{center}
\caption{(Color online) Profiles of pressure (left) and density (right) in a power-law trap with $\alpha=0.5$ (top) and $\alpha=2$ (bottom) at temperatures $\hat{T}=0.02, 0.5, 1$ for rods of sizes $\sigma=1$ (solid curves) and $\sigma=\infty$ (dashed curves).}
  \label{fig:fig3}
\end{figure}

In Fig.~\ref{fig:fig3} we show pressure and density profiles for the two extreme rod sizes in two different power-law potentials.
The shape of one potential $(\alpha=\frac{1}{2})$ is concave and that of the other $(\alpha=2)$ convex.
Corresponding profiles for a linear potential $(\alpha=1)$ have already been shown in Fig.~\ref{fig:fig1} albeit on a somewhat different scale.

The opposite $\hat{T}$-dependence of the pressure near the center of the trap is evident. 
The curves at the lowest temperature are close to the $\hat{T}=0$ profiles (\ref{eq:50}).
Naturally, the pressure decrease with rising $\hat{T}$ in for $\alpha=\frac{1}{2}$ has a much larger effect on the density than does the pressure increase for $\alpha=2$.
Another noteworthy feature is that the difference in profile between the lattice system and the continuum system is much more pronounced for the convex potential than for the concave potential.

\subsection{Oscillations in density profile}\label{sec:powe-soft-wall}
In Sec.~\ref{sec:grav-dft} we have already identified some hard-floor effects in the form of spatially attenuated oscillations in the density profiles $\bar{n}_i$ and $\rho_i$ of rods and mass, respectively.
We found (in Fig.~\ref{fig:fig9}) that the effect is very conspicuous in the former and partially averaged out in the latter.
A more systematic analysis of hard-wall effects for rods of various sizes on the lattice and for rods in a continuum will be presented in Sec.~\ref{sec:ster}.

\begin{figure}[b]
  \begin{center}
  \includegraphics[width=43mm]{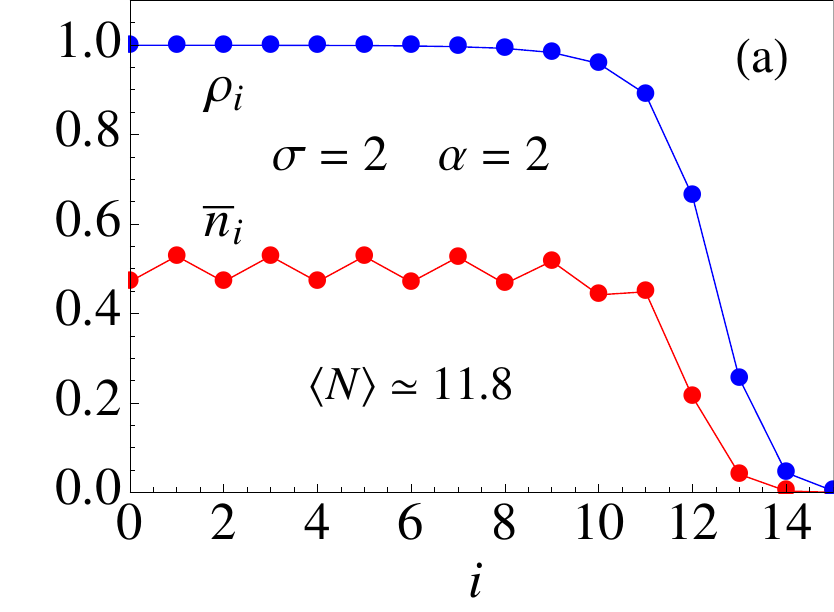}%
 \includegraphics[width=43mm]{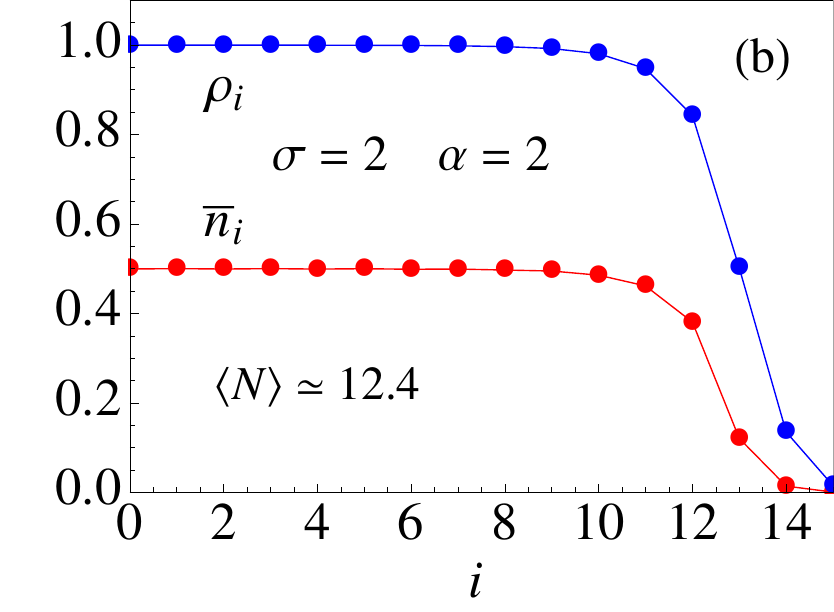} 
  \includegraphics[width=43mm]{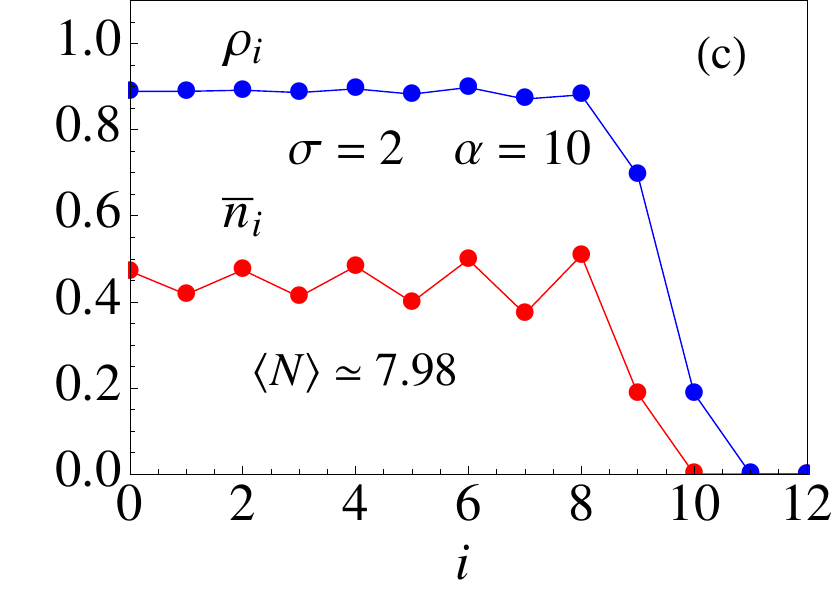}%
 \includegraphics[width=43mm]{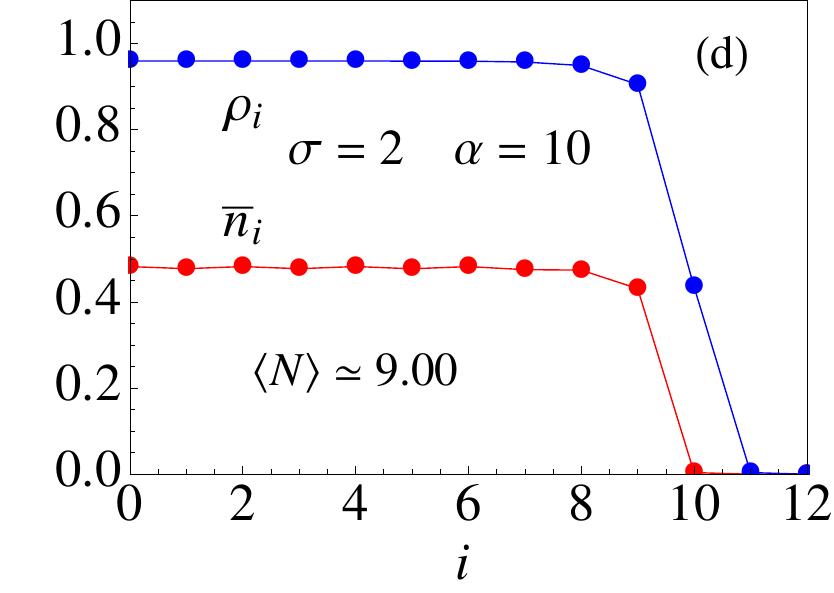} 
\end{center}
\caption{(Color online) Density profile of rods $\bar{n}_i$ and mass $\rho_i$ for rods of size $\sigma=2$ in a harmonic trap $(\alpha=2)$ and in a power-law trap with firmer walls $(\alpha=10)$. In both traps we use $x_0=10V_\mathrm{c}$. 
All data are for $\bar{T}=0.1$. The chemical potential $\bar{\mu}$ has the values (a) 1.0, (b) 1.55, (c) 0.3, (d) 0.5.}
  \label{fig:fig15}
\end{figure}

Here we briefly examine the question whether the soft walls of power-law traps also produce patterns of spatial density oscillations.
We begin with rods of size $\sigma=2$.
In Fig.~\ref{fig:fig15} we compare data for traps with walls of two different degrees of softness.
These data are produced by using the potential (\ref{eq:26}) with $x=iV_\mathrm{c}$ and calculating the profiles from (\ref{eq:b10}) and (\ref{eq:101}). We use $M=2I_\mathrm{max}+1$ with $I_\mathrm{max}=20$ and a shift that positions the rod with index $i=0$ at the center of the trap.

The harmonic trap is, in some sense, the smoothest form of confinement.
We see in panels (a) and (b) that oscillations do make their appearance in the $\bar{n}_i$-profiles. 
The amplitudes of these oscillations tend to be rather uniform across the region where the rod population is significant.
The general trend is that the amplitudes increase with increasing chemical potential $\bar{\mu}\doteq\mu/u_0$ or decreasing temperature $\bar{T}\doteq k_BT/u_0$.

Superimposed on this systematic trend is an oscillatory dependence on $\bar{\mu}$. 
The data in the panels (a) and (b) are for two successive values near maximum and minimum amplitude of $\bar{n}_i$-oscillations. 
The minimum amplitude is almost imperceptible.
We also note that in the $\rho_i$-profiles these oscillations are almost completely averaged out.

Increasing the stiffness of the trap walls produces no dramatic changes. 
The dependence on $\bar{\mu}$ of the spatial oscillations remains qualitatively similar. 
The amplitudes still oscillate as $\bar{\mu}$ is varied.
The data in the panels (c) and (d) are for $\alpha=10$ and for values of $\bar{\mu}$ near successive maximum and minimum amplitude. 

One systematic trend as the trap wall becomes increasingly firm is that the spatial oscillations become weaker near the center of the trap and stronger near the walls.
This trend is visible between the data in panels (a) and (c), for example.
We shall see in Sec.~\ref{sec:ster} that the oscillatory dependence of the amplitudes on $\bar{\mu}$ disappears when the walls become hard $(\alpha\to\infty)$.

\begin{figure}[t]
  \begin{center}
  \includegraphics[width=43mm]{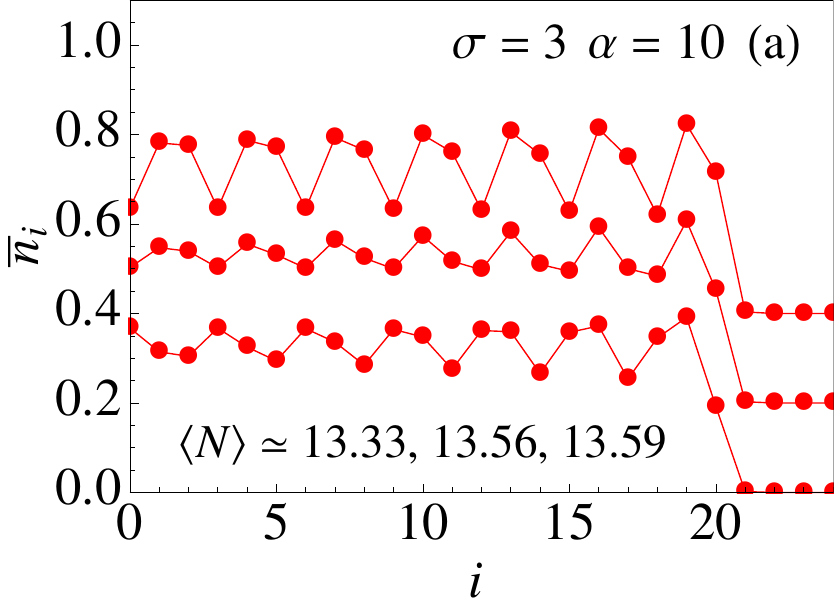}%
 \includegraphics[width=43mm]{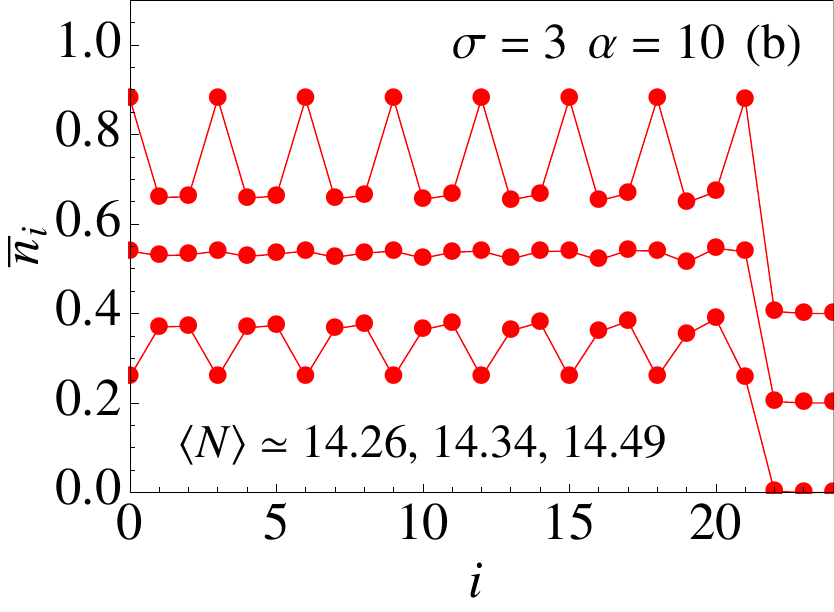} 
\end{center}
\caption{(Color online) Density profile $\bar{n}_i$ for rods of size $\sigma=3$ in a power-law trap with $\alpha=10$. We use $x_0=20V_\mathrm{c}$. All data are for $\bar{T}=0.1$. The scaled chemical potentials  are (a) $\bar{\mu}=1.1, 1.15, 1.2$ and (b) $\bar{\mu}=2.1, 2.14, 2.2$. In each plot the second and third data sets are vertically displaced by 0.2 and 0.4, respectively.}
  \label{fig:fig16}
\end{figure}

In Fig.~\ref{fig:fig16} we show some data for rods of size $\sigma=3$ in a trap with relatively stiff walls $(\alpha=10)$ and twice the width of the one used before. 
The calculations are based on Eqs.~(\ref{eq:b15}) and (\ref{eq:b16}).
Here we only show data of the $\bar{n}_i$-profiles for two conditions.

The average numbers of rods in the data of panel (a) are such that the rods easily fit into the space at low potential energy. 
We observe significant oscillations with a period near but not exactly three. 
A small change in chemical potential produces shifts in the oscillatory patterns and variations in the average amplitude.
We also observe slightly larger amplitudes near the wall compared to the center of the trap.
At significantly smaller values of $\langle N\rangle$ the oscillations are much weaker.

The data in panel (b) are for circumstances where the rods are squeezed into the trap. 
Here the rod positions are more correlated. 
The oscillations are closer to period three.
We also observe the oscillatory dependence of the amplitude on $\bar{\mu}$.
For some values there are one or two dominant configurations, producing high amplitude.
For other values, there are three configuration that have very similar statistical weight, producing low amplitude.

%
\section{Steric wall effects}\label{sec:ster}
%
Microscopic density profiles of rods or other particles with shapes near hard walls are relevant in the contexts of granular matter, porous solids, and zeolites among others.
Robledo and Rowlinson \cite{Robledo/Rowlinson:1986} studied the effects of confinement on hard rods in a continuum.
Davis \cite{Davis:1990} extended that study to include first-neighbor interactions between rods.
More recently, Ibsen {\it et al} \cite{Ibsen/etal:1997} reported a general and exact solution for hard rods confined by a gravitational field and a hard floor.
A computer simulation study of Mehrotra {\it et al} \cite{Mehrotra/etal:2011} of hard spheres under the same confinement in 3D produces similar results.

Here we pick up threads from Secs.~\ref{sec:grav-dft} and ~\ref{sec:powe-soft-wall} to investigate steric wall effects of hard rods on a lattice.
One goal is to showcase the versatility of the method of analysis presented in Appendix~\ref{sec:appb} and to establish how it connects to the continuum analysis familiar from previous work.

We consider a box with rigid walls and investigate the oscillations in the density profiles produced by the steric interactions between rods of size $\sigma\geq2$.
The effects of a single wall, relevant in sufficiently wide boxes, can be determined analytically for rods of any size on the lattice and for rods in the continuum. 
Two walls within the distance of a certain coherence length affect the density profile from opposite sides.
That coherence length is shown to grow with average density.
We combine exact analytic results with results from a rigorous recursive scheme.

\subsection{$\sigma=2$}\label{ster-sig-2}
We begin with the case of a semi-infinite box with one wall at $i=I$ and the other at $i\to-\infty$  populated by rods of size $\sigma=2$ to an average mass density $0<\rho^{(\mathrm{mes})}<1$. 
Later we move the second wall to $i=-I$.
For the semi-infinite box we have found an analytic solution.
The result (with $j=I-i$ in the present context) turns out to have a simple structure:
\begin{equation}\label{eq:103} 
\bar{n}_i=\frac{1}{2}\rho^{(\mathrm{mes})}\left[1-\left(\frac{\rho^{(\mathrm{mes})}}{\rho^{(\mathrm{mes})}-2}\right)^j\,\right],\quad
j=0,1,2,\ldots
\end{equation}
The relation between the fugacity and the average mass density is
\begin{equation}\label{eq:104} 
\rho^{(\mathrm{mes})}=1-\frac{1}{\sqrt{1+4\zeta}} ~~\mathrm{or}~~ \zeta=\frac{1}{4}\left[\frac{1}{(1-\rho^{(\mathrm{mes})})^2}-1\right].
\end{equation}
We have derived $\bar{n}_1$ in (\ref{eq:103}) directly from (\ref{eq:b11}) with $g_2$ in the form of an infinite continued fraction that is readily evaluated.
The $\bar{n}_i$ for $i=2,3,\ldots$ then follow directly from (\ref{eq:b5}).
The oscillations in the probability distribution $\bar{n}_i$ of rod positions thus decay exponentially with distance from the wall.
The boundary coherence length,
\begin{equation}\label{eq:106} 
\xi=\frac{1}{\ln(2/\rho^{(\mathrm{mes})}-1)},
\end{equation}
vanishes for $\rho^{(\mathrm{mes})}\to0$ and diverges for $\rho^{(\mathrm{mes})}\to1$.

\begin{figure}[t]
  \begin{center}
 \includegraphics[width=43mm]{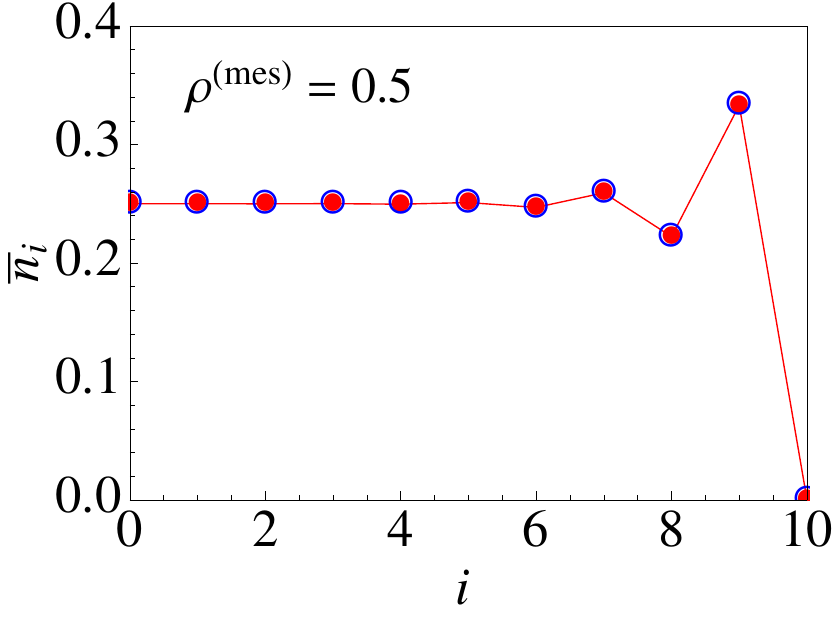}%
 \includegraphics[width=43mm]{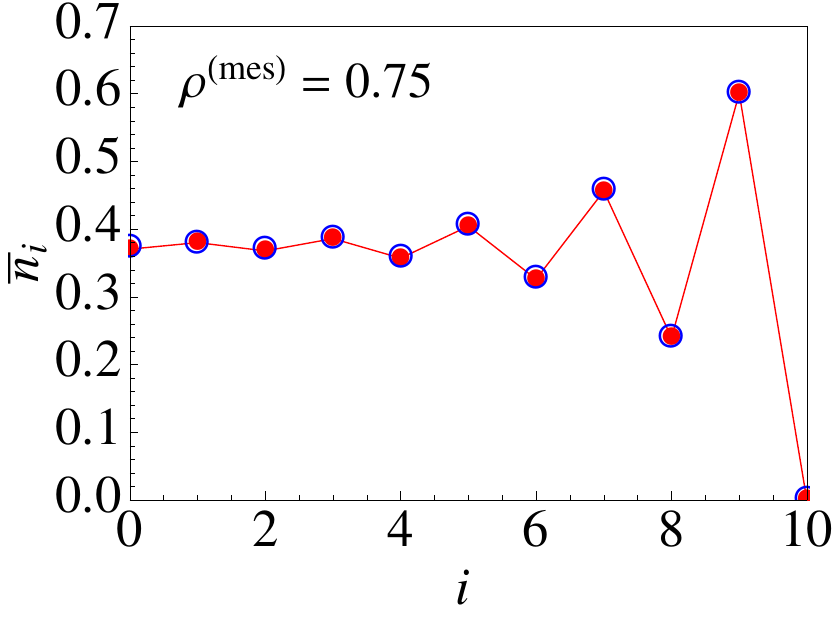}
  \includegraphics[width=43mm]{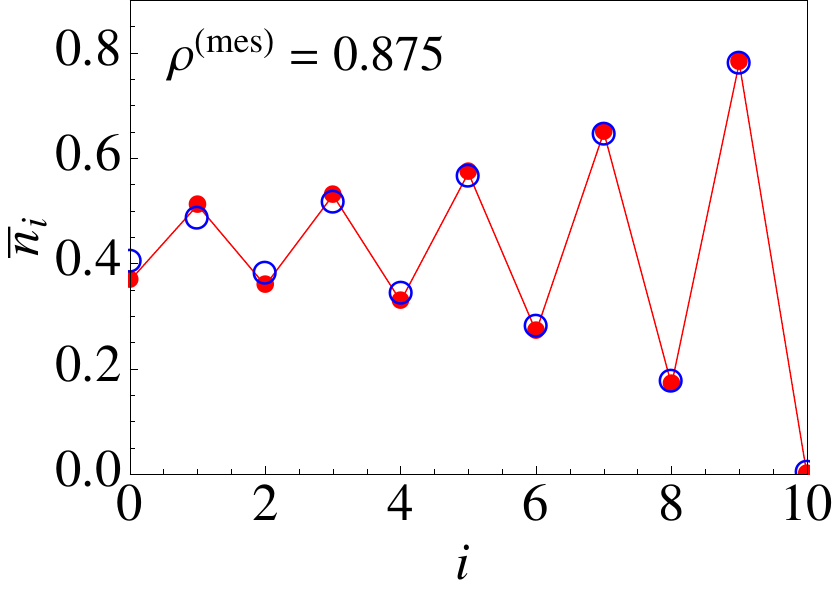}%
 \includegraphics[width=43mm]{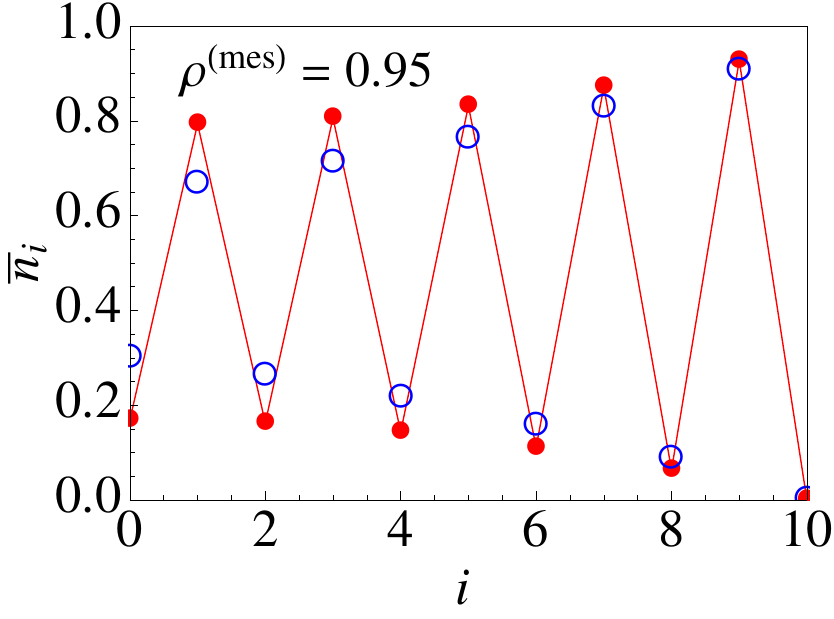}
\end{center}
\caption{(Color online) Density profile for rods of size $\sigma=2$ with center at coordinate $i=0,\pm1,\ldots,I$ in a box of width $I=10$. 
Full circles connected by lines: solution of (\ref{eq:b10}). Open circles: result (\ref{eq:103}) for a box of infinite width with one wall at $i=10$.}
  \label{fig:fig12}
\end{figure}

Next we examine how the oscillations near the wall at $i=I$ are affected by the presence of a second wall at $i=-I$.
For that purpose we have solved (\ref{eq:b10}) with $\zeta$ from (\ref{eq:104}) for comparison with the analytic solution (\ref{eq:103}).
The results, shown in Fig.~\ref{fig:fig12}, are almost indistinguishable for $\rho^{(\mathrm{mes})}\lesssim0.75$.
Here the coherence length (\ref{eq:106}) is much smaller than the distance between the walls.
At larger mass density the two sets of results begin to deviate from each other as $\xi$ grows and reaches about half the wall-to-wall distance at $\rho^{(\mathrm{mes})}\lesssim0.95$.

The signature wall effect for $\sigma=2$ manifests itself in the form of attenuated, period-2 spatial oscillations in the $\bar{n}_i$. 
The exponential attenuation is governed by a coherence length that grows with the density of rods in the box. 
In the limit $\rho^{(\mathrm{mes})}\to1$ the oscillations persist across the box as expected.

\subsection{$2<\sigma<\infty$}\label{ster-sig-3p}
We continue with the analysis of a system of rods of size $2<\sigma<\infty$ subject to the potential
\begin{equation}\label{eq:114} 
\mathcal{U}_i=\left\{\begin{array}{ll} \infty &:~ i\leq0,  \rule[-2mm]{0mm}{5mm} \\ 0 &:~ i>0,
 \rule[-2mm]{0mm}{5mm} \end{array} \right.
\end{equation}
representing a single rigid wall at $i=0$.
We present the exact solution of Eqs.~(\ref{eq:b2}) adapted to this case,
\begin{equation}\label{eq:115} 
\zeta=\bar{n}_{i}\!\!\prod\limits_{k=i}^{i+\sigma-2}
\left[1-\!\!\!\sum\limits_{j=k-\sigma+2}^{k}\bar{n}_{j}\right]
\prod\limits_{k=i}^{i+\sigma-1}\left[1-\!\!\!\sum\limits_{j=k-\sigma+1}^{k}\bar{n}_{j}\right]^{-1}\!\!\!.
\end{equation}
We have $\bar{n}_i=0$ for $i\leq0$ and assume that far from the wall $(i\gg1)$ the $\bar{n}_i$ approach uniformity,
\begin{equation}\label{eq:116} 
\lim_{i\to\infty}\bar{n}_i=\bar{n}^{(\mathrm{mes})}=\frac{\rho^{(\mathrm{mes})}}{\sigma},
\end{equation}
where $0<\rho^{(\mathrm{mes})}<1$ is the mass density in the bulk (average cell occupancy).
The fugacity, the only control variable aside from $\sigma$, depends on the asymptotic solution (\ref{eq:116}) as follows:
\begin{equation}\label{eq:117} 
\zeta=\frac{\bar{n}^{(\mathrm{mes})}\big[1-(\sigma-1)\bar{n}^{(\mathrm{mes})}\big]^{\sigma-1}}
{\big[1-\sigma \bar{n}^{(\mathrm{mes})}\big]^\sigma}, \quad 0<\bar{n}^{(\mathrm{mes})}<\frac{1}{\sigma}.
\end{equation}
Next  we convert (\ref{eq:115}) into a recursion relation that expresses the solution at a given site as a function of the solutions at the $\sigma-1$ sites immediately to its left:
\begin{align}\label{eq:118} 
\bar{n}_{i} &=1-\sum_{j=i-\sigma+1}^{i-1} \bar{n}_j \nonumber \\
&-\frac{\bar{n}_{i-\sigma+1}}{\zeta}\prod\limits_{k=i-\sigma+1}^{i-1}
\left[\frac{1-\sum\limits_{j=k-\sigma+2}^{k}\bar{n}_{j}}{1-\sum\limits_{j=k-\sigma+1}^{k}\bar{n}_{j}}\right],
\end{align}
for $i=\sigma,\sigma+1,\ldots$ and with $\zeta$ from (\ref{eq:117}).
This recursion relation depends on the $\sigma-1$ parameters $\bar{n}_1,\ldots,\bar{n}_{\sigma-1}$.
Assuming that the solution with asymptotics (\ref{eq:116}) is unique, these parameters can be found with a little guidance from (\ref{eq:103}) for the case $\sigma=2$.
They are
\begin{equation}\label{eq:119} 
\bar{n}_i=\bar{n}^{(\mathrm{mes})}\frac{\big[1-\sigma \bar{n}^{(\mathrm{mes})}\big]^{i-1}}
{\big[1-(\sigma-1)\bar{n}^{(\mathrm{mes})}\big]^i},\quad i=1,\ldots,\sigma-1.
\end{equation}

\begin{figure}[b]
  \begin{center}
  \includegraphics[width=43mm]{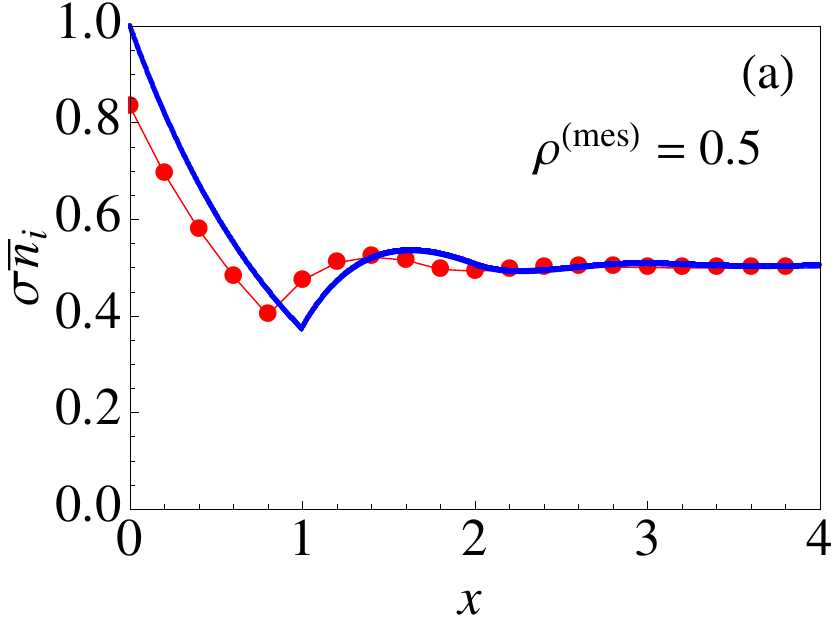}%
 \includegraphics[width=43mm]{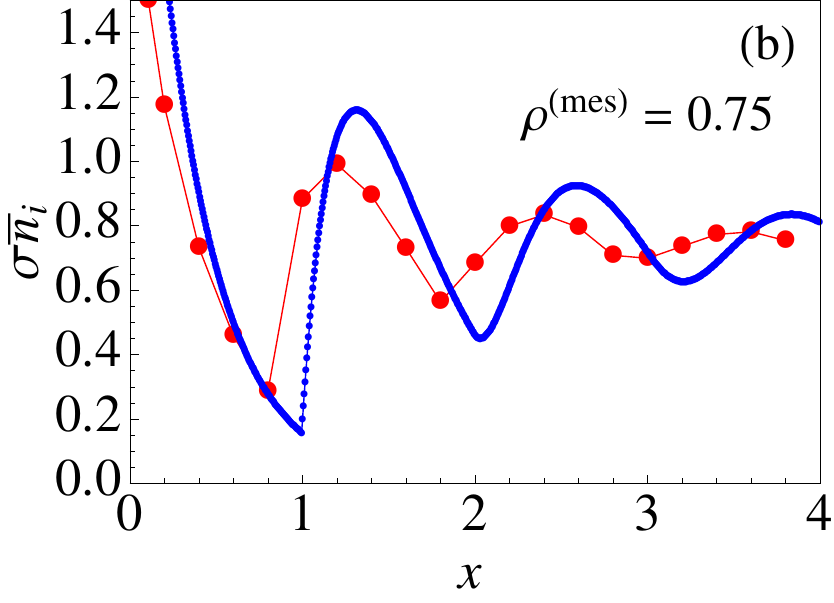} 
\end{center}
\caption{(Color online) Scaled density of rods of size $\sigma=5$ (big circles) and $\sigma=200$ (small circles) with first cell at position $x_i=(i-1)/\sigma$ near a rigid wall at $i=0$.}
  \label{fig:figa1}
\end{figure}

In Fig.~\ref{fig:figa1} we show data generated recursively from (\ref{eq:118}) for the scaled density $\bar{\nu}_i\doteq\sigma \bar{n}_i$ versus the scaled position $x_i\doteq(i-1)/\sigma$ for rods of sizes $\sigma=5$ and $\sigma=200$.
The latter size is meant to generate an impression of what to expect in the continuum limit.
The data suggest that the dominant wall effect manifests itself again in an attenuated spatial oscillation.
Successive minima are approximately spaced by $\sigma$.
Only in the limit ${\rho^{(\mathrm{mes})}\to1}$, when the attenuation disappears, are the oscillations locked into the wavelength $\sigma$.
This raises the interesting question, best analyzed in the continuum limit, what the spectrum of the wall oscillations is and how it depends on $\rho^{(\mathrm{mes})}$.

\subsection{$\sigma=\infty$}\label{ster-sig-in}
The continuum limit carried out for $0<x<1$ produces an exponential function as follows:
\begin{equation}\label{eq:120} 
\bar{\nu}(x)=se^{-sx},\quad s\doteq\frac{\rho^{(\mathrm{mes})}}{1-\rho^{(\mathrm{mes})}}.
\end{equation}
This result can now be extended to $x>1$ by using a continuum version of the fugacity (\ref{eq:117}),
\begin{equation}\label{eq:121} 
\zeta=se^s,
\end{equation}
and a continuum version of (\ref{eq:115}),
\begin{equation}\label{eq:122} 
h(x)=\zeta\exp\left(-\int_{x-1}^xdx'\,h(x')\right),
\end{equation}
where 
\begin{equation}\label{eq:a123} 
h(x)\doteq\frac{\bar{\nu}(x)}{1-\int_{x}^{x+1}dx'\,\bar{\nu}(x')},
\end{equation}
as derived Percus \cite{Percus:1976}.
Following Vanderlick et al. \cite{Vanderlick/etal:1986}, we convert (\ref{eq:122}) and (\ref{eq:a123}) into difference-differential equation,
\begin{equation}\label{eq:124} 
\frac{d}{dx}h(x)=h(x)\big[h(x-1)-h(x)\big],
\end{equation}
\begin{equation}\label{eq:125} 
\frac{d}{dx}\left[\frac{\bar{\nu}(x)}{h(x)}\right]=\bar{\nu}(x)-\bar{\nu}(x+1),
\end{equation}
respectively.
Next we integrate (\ref{eq:124}) and (\ref{eq:125}) in alternating sequence over intervals of unit length, using (\ref{eq:120}) and $h(x)\equiv0$ for $x<0$.
We thus obtain the following exact, continuous, and piecewise analytic expression for $\bar{\nu}$ on successive intervals $m<x<m+1$:
\begin{equation}\label{eq:126} 
\bar{\nu}(x)=\sum_{k=0}^m\frac{s^{k+1}}{k!}(x-k)^ke^{-(x-k)s}
\end{equation}
with asymptotic value,
\begin{equation}\label{eq:150} 
\lim_{x\to\infty}\bar{\nu}(x)=\frac{s}{1+s}=\rho^{(\mathrm{mes})},
\end{equation}far from the wall, approached more and more slowly with increasing average mass density $\rho^{(\mathrm{mes})}$.
The lattice DFT analysis thus connects neatly with known results \cite{Davis:1990, Ibsen/etal:1997, Robledo/Rowlinson:1986, Percus:1976, Vanderlick/etal:1986} for monodisperse rods in a continuum.

\begin{figure}[htb]
  \begin{center}
  \includegraphics[width=43mm]{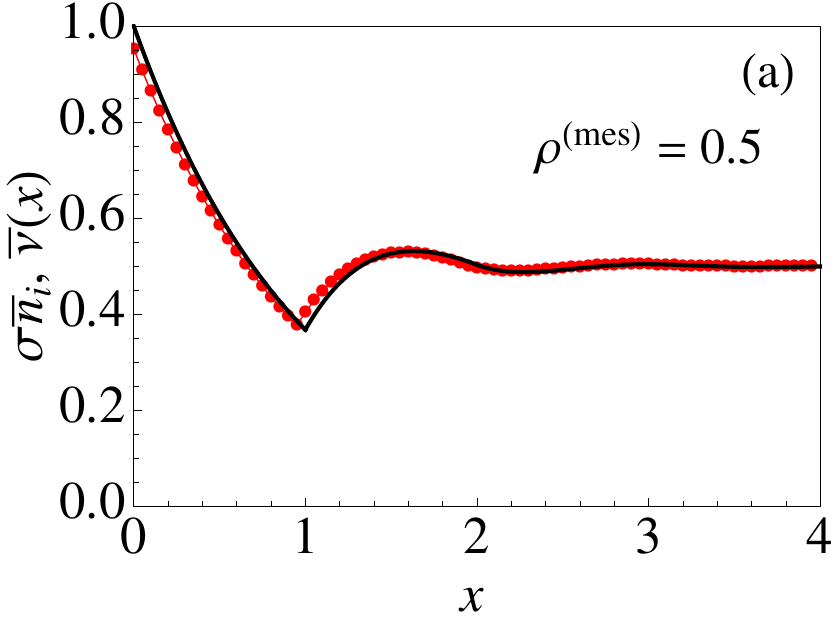}%
 \includegraphics[width=43mm]{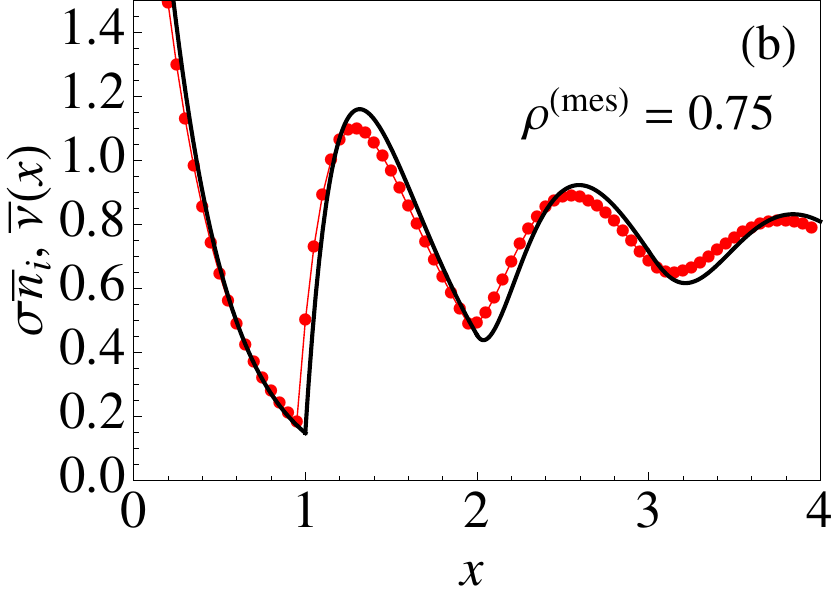} 
\end{center}
\caption{(Color online) Scaled density of rods of size $\sigma=20$ on a lattice (big circles) and rods of (scaled) unit size in a continuum (solid lines).}
  \label{fig:figa2}
\end{figure}

In Fig.~\ref{fig:figa2} we graphically compare this analytic solution (\ref{eq:126}) for the continuum model with the iterative solution (\ref{eq:118}) for the lattice model with $\sigma=20$.
The dominant feature of the curves is an attenuated spatial oscillation.
The singularities at $x=m$ become progressively weaker with increasing $m$: 
$d^k\bar{\nu}(x)/dx^k$ is continuous at $x=m$ for $k<m$.
Except for $m=1$, the singularities do not coincide with the minima of $\bar{\nu}(x)$.

\begin{figure}[b]
  \begin{center}
  \includegraphics[width=43mm]{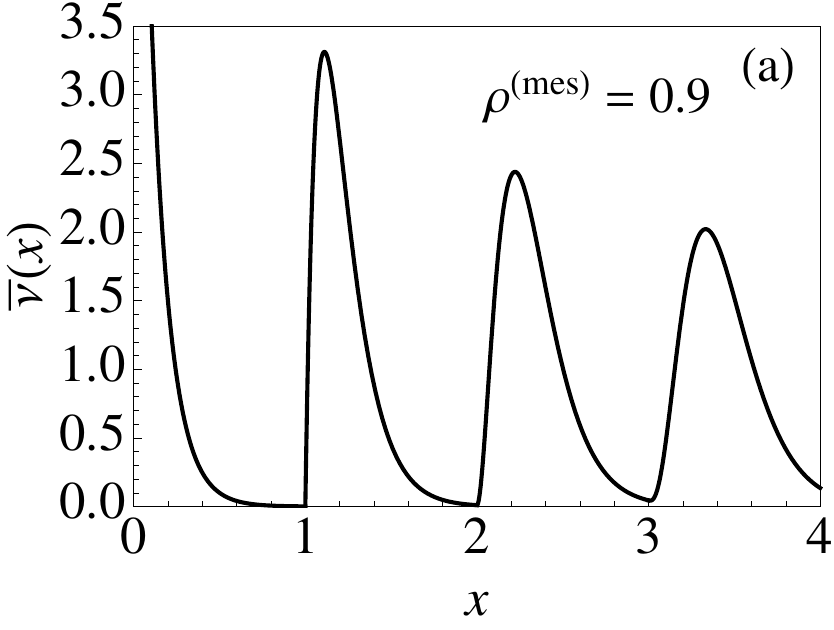}%
 \includegraphics[width=43mm]{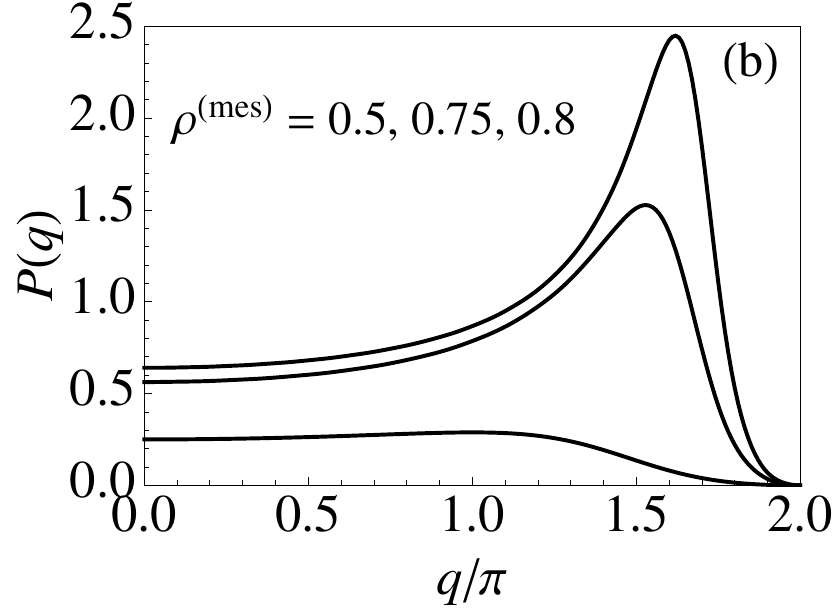} 
\end{center}
\caption{(a) Scaled density of rods in a continuum with $\rho^{(\mathrm{mes})}=0.9$. (b) Spectrum (\ref{eq:152}) of the attenuated oscillation (\ref{eq:126}) at three values of  $\rho^{(\mathrm{mes})}$}
  \label{fig:fig17}
\end{figure}

The limit $\rho^{(\mathrm{mes})}\to1$ $(s\to\infty)$ is subtle. 
As the population of rods becomes more crowded, they begin to line up with increasing probability near the integer positions as illustrated in Fig.~\ref{fig:fig17}(a).
In the limit $\rho^{(\mathrm{mes})}\to1$, maxima and minima approach the singularity values $x=m$ in pairs from opposite sides. The maxima diverge and the minima approach zero.
The area under the curve between successive singularities approaches unity.
The shape of the curve approaches an \textsf{L} of infinite height and unit width, effectively the function
\begin{equation}\label{eq:151} 
\lim_{s\to\infty}\bar{\nu}(x)=\sum_{m=0}^\infty\delta(x-m).
\end{equation}

In Fig.~\ref{fig:fig17}(b) we show the spectrum of the attenuated oscillation (\ref{eq:126}),
\begin{equation}\label{eq:152} 
\bar{P}(q)\doteq 2\int_0^\infty dx \left[\bar{\nu}(x)-\rho^{(\mathrm{mes})}\right]\cos(qx).
\end{equation}
This quantity features a peak at wave number ${1\lesssim q/\pi<2}$ emerging from a broad and flat background. 
The Fourier transform (\ref{eq:152}) and the limit $s\to\infty$ are not interchangeable operations.
With increasing $\rho^{(\mathrm{mes})}$ the peak becomes taller and sharper as it moves toward commensurability at $q/\pi=2$.

\begin{figure}[htb]
  \begin{center}
  \includegraphics[width=43mm]{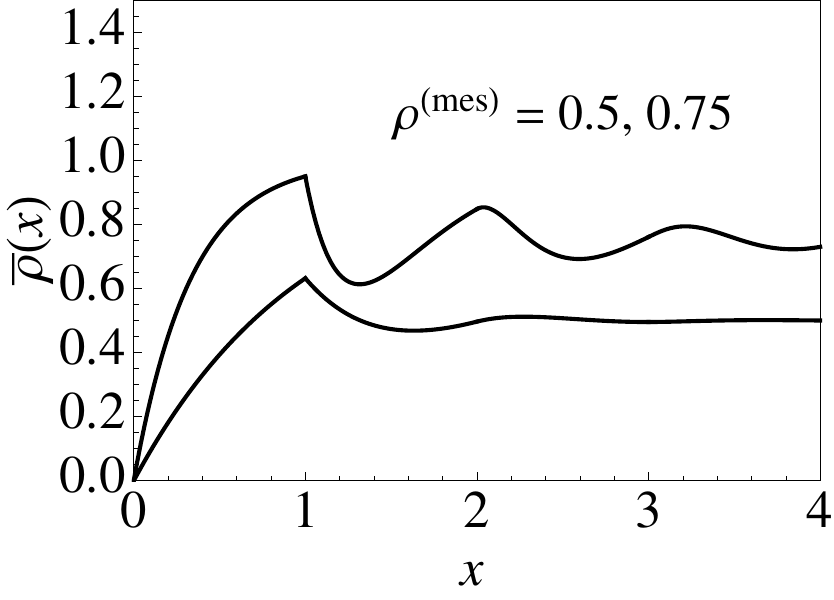}%
 \includegraphics[width=43mm]{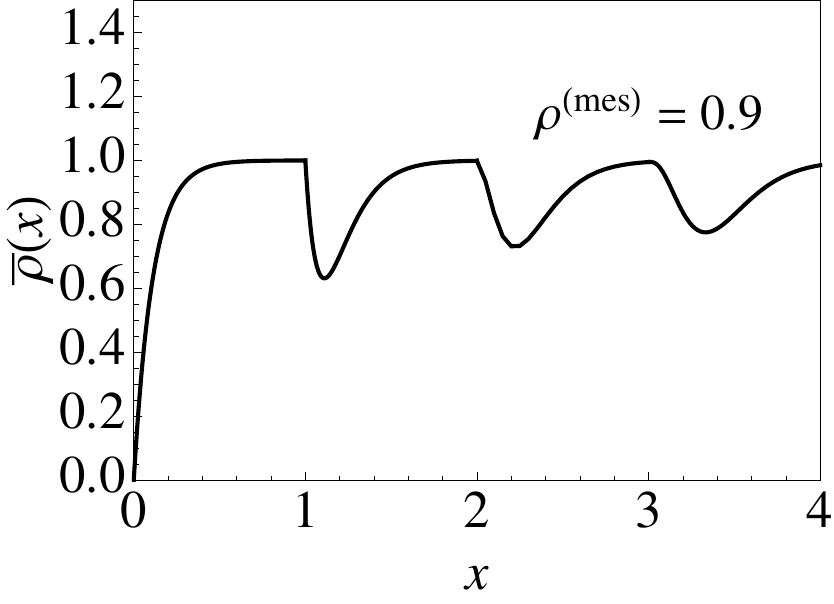} 
\end{center}
\caption{Profiles of mass density of rods in a continuum at three different average values $\rho^{(\mathrm{mes})}$.}
  \label{fig:fig18}
\end{figure}

The DFT mass density in the continuum, calculated from (\ref{eq:126}) via a continuum version of (\ref{eq:101}),
\begin{equation}\label{eq:156} 
\bar{\rho}(x)\doteq \int_{x-1}^x dx'\bar{\nu}(x'),
\end{equation}
becomes
\begin{equation}\label{eq:157} 
\bar{\rho}(x)=1-\sum_{k=0}^m\frac{s^k}{k!}(x-k)^ke^{-(x-k)s}
\end{equation}
on successive intervals $m<x<m+1$.
In Fig.~\ref{fig:fig18} we show profiles of $\bar{\rho}(x)$ for the corresponding to the profiles of $\bar{\nu}(x)$ shown in Figs.~\ref{fig:figa2} and \ref{fig:fig17}.
The oscillatory wall effects are somewhat milder in $\bar{\rho}(x)$ than in $\bar{\nu}(x)$ but still fairly conspicuous.
In the limit $s\to\infty$ the oscillations in $\bar{\nu}(x)$ diverge as in (\ref{eq:151}) but disappear in $\bar{\rho}(x)$, which becomes constant.

%
\section{Conclusions and outlook}\label{sec:conc}
%
In this work we have investigated the equilibrium statistical mechanics of monodisperse hard rods confined by external potentials, producing characteristic spatial profiles of density and pressure.
The rods occupy $\sigma$ consecutive cells of volume $V_\mathrm{c}$ in a linear array.
The continuum limit is implemented as $\sigma\to\infty$, $V_\mathrm{c}\to0$ with $\sigma V_\mathrm{c}$ kept finite and nonzero.
The rods interact via hardcore exclusion forces. 
Contact forces have also been considered.

We have been employing two distinct methods of analysis, density functionals and generalized exclusion statistics, with partially overlapping domains of applicability and somewhat complementary strengths.
The usefulness, soundness, and consistency of the two approaches has been demonstrated in a series of applications that include confinement by rigid walls, gravity, and power-law traps.

We have shown that profiles of density, for example, exhibit characteristic features on a mesoscopic length scale that are reproduced identically by both approaches and, for $\sigma\geq2$, additional features on a microscopic length scale that are only resolved in the DFT analysis.
The DFT pressure profiles are also compared with profiles of average microscopic pressure inferred from pair distribution functions in the context of one application.
Finally, we have established contact between our results for the lattice system and prior results for corresponding continuum systems, demonstrating full consistency

The statistical mechanical analysis of hard rods of mixed sizes presents itself as a natural and promising extension of this work.
Significant ground has already been broken via both calculational tools used in this work.
The mathematical structure of exact density functionals for hard-rod mixtures has recently been established  \cite{Bakhti/etal:2012} and is awaiting analysis in specific applications.

Hard rod ensembles of mixed sizes on a lattice are mathematically equivalent to magnetic domains in Ising chains.
The combinatorics and statistical mechanics of such domains analyzed via generalized exclusion statistics have been in place for some time \cite{Lu/etal:2008} and is awaiting extensions to heterogeneous environments and new applications of current interest.
The completion of the work reported here has paved the way for projects along these lines \cite{poharo}.

\appendix

%
\section{Local pressure from momentum flow}\label{sec:appa}
%
In terms of the particle (hard rod) momenta $\Pi_n(t)$ and positions $x_n(t)$, the microscopic momentum density in Eq.~(\ref{eq:momentum-flow}) is given by $\Pi(x,t)=\sum_n \Pi_n(t)\delta(x-x_n(t))$ . 
The time derivative of its Fourier transform $\hat\Pi(k,t)=\int\mathrm{d}x\,\Pi(x,t)\,e^{-ikx}\, =\sum_n \Pi_n(t)\,e^{-ikx_n(t)}$ reads (with $m_\mathrm{r}$ the mass of the rods, and assuming only pair interaction forces to be present)
\begin{align}\label{eq:dPidt}
\frac{\partial\hat\Pi}{\partial t}=&
-ik\sum_n \frac{\Pi_n(t)^2}{m_\mathrm{r}}\,e^{-ikx_n(t)}\\
&+\sum_{n,m} f_{mn}(t)\,e^{-ikx_n(t)}+\sum_{n}f_n^\mathrm{ext}(t)\,e^{-ikx_n(t)}\,.
\nonumber\end{align}
Here $f_{mn}(t)\equiv f(x_m(t),x_n(t))$ is the force of the $m$th on the $n$th particle ($f_{nn}\equiv0$), $f_n^\mathrm{ext}(t)=f^\mathrm{ext}(x_n(t))$ the external force on the $n$th particle, and we have used $\dot x_n=\Pi_n/m_\mathrm{r}$ and Newton's equations $\dot\Pi_n= f_n^\mathrm{ext}+\sum_{m} f_{nm}$.
For interaction forces obeying the principle of action and reaction, $f_{mn}=-f_{nm}$, the second term on the right hand side of Eq.~(\ref{eq:dPidt}) can be written in the form $(1/2)\sum_{n,m} f_{mn}\,(e^{-ikx_n}-e^{-ikx_m})= -ik\sum_{m,n} f_{mn}x_{mn}e^{-ikX_{mn}}[\sin(kx_{mn})/(kx_{mn})]$, where $X_{mn}\equiv(x_m+x_n)/2$ and $x_{mn}\equiv (x_n-x_m)/2$.
Comparing Eq.~(\ref{eq:dPidt}) with the Fourier-transformed right hand side of Eq.~(\ref{eq:momentum-flow}), the microscopic pressure in Fourier space becomes
\begin{align}\label{eq:pmic-fourier}
\hat p_\mathrm{mic}(k,t)=&\sum_n \frac{\Pi_n(t)^2}{m_\mathrm{r}}\,e^{-ikx_n(t)}\\
&+\sum_{m,n} f_{mn}(t)x_{mn}(t)\,\frac{\sin(kx_{mn}(t))}{kx_{mn}(t)}e^{-ikX_{mn}(t)}\,.\nonumber
\end{align}
To obtain the corresponding expression in real space, we calculate
\begin{align}\label{eq:back-transform}
\int\frac{\mathrm{d}k}{2\pi}&e^{-ikX_{mn}+ikx}\frac{\sin(kx_{mn})}{kx_{mn}}\hspace*{10em}\nonumber\\
&=\frac{1}{2}\int\frac{\mathrm{d}k}{2\pi}e^{-ikX_{mn}+ikx}\int_{-1}^1\mathrm{d}\alpha\,e^{ikx_{mn}\alpha}\nonumber\\
&=\frac{1}{2}\int_{-1}^1\mathrm{d}\alpha\,\delta(x-X_{mn}+\alpha x_{mn})\nonumber\\
&=\frac{1}{2|x_{mn}|}\,g(x;x_m,x_n)\,,
\end{align}
where
\begin{align}\label{eq:g-function}
g(x;x_m,x_n)&=\left\{\begin{array}{ll}
1\,, & x\in[\min(x_m,x_n),\max(x_m,x_n)]\,,\\[1ex]
0\,, & \textrm{else}\,.\end{array}\right.
\end{align}
Accordingly,
\begin{align}\label{eq:pmic}
p_\mathrm{mic}(x,t)=&\sum_n \frac{\Pi_n(t)^2}{m_\mathrm{r}}\delta(x-x_n(t))\\
&+\frac{1}{2}\sum_{m,n} f_{mn}(t)\frac{x_{mn}(t)}{|x_{mn}(t)|}\,g(x;x_m(t),x_n(t))\,.\nonumber
\end{align}
Using $f_{mn}x_{mn}=f_{nm}x_{nm}$ and Eq.~(\ref{eq:g-function}), the double sum can be written as $\sum_{x_m<x}\sum_{x_n>x} f_{mn}x_{mn}/|x_{mn}|$, and the equilibrium average of Eq.~(\ref{eq:pmic}) then yields Eq.~(\ref{eq:p-rho2}).

%
\section{Density profiles}\label{sec:appb}
%
Consider a system of rods  of size $\sigma$ confined to a finite array of cells numbered $1,\ldots,L=M+\sigma-1$.
The hardcore repulsion is the only interaction between rods.
The external potential $\mathcal{U}_i$ is arbitrary.
The free energy functional (\ref{eq:freeenergy-dft}) simplifies into 
\begin{align}\label{eq:b1}
\beta F=&
\sum_{i=1}^{L}\Biggl\{\tilde{n}_i\ln \tilde{n}_i+\Bigl(1\!-\!\sum_{j=i-\sigma+1}^i\tilde{n}_j\Bigr)\ln\Bigl(1\!-\!\sum_{j=i-\sigma+1}^i\tilde{n}_j\Bigr)\nonumber\\
&-\Bigl(1\!-\!\sum_{j=i-\sigma+1}^{i-1}\tilde{n}_j \Bigr)\ln\Bigl(1\!-\!\sum_{j=i-\sigma+1}^{i-1}\tilde{n}_j \Bigr) \Biggr\},
\end{align}
where $\tilde{n}_1,\ldots,\tilde{n}_M$ are the (yet undetermined) probabilities of rods at the allowed positions.
The minimization (\ref{eq:structure-equations}) then leads to the following relations that determine the exact density profile of rods, $\{\bar{n}_i\}$, for an any given external potential $\mathcal{U}_i$  at given temperature $T$ and chemical potential $\mu$:
\begin{equation}\label{eq:b2} 
e^{-\beta(\mathcal{U}_i-\mu)}=\bar{n}_{i}\,\frac{\prod\limits_{k=i}^{i+\sigma-2}
\left[1-\sum\limits_{j=k-\sigma+2}^{k}\bar{n}_{j}\right]}{\prod\limits_{k=i}^{i+\sigma-1}\left[1-\sum\limits_{j=k-\sigma+1}^{k}\bar{n}_{j}\right]},
\end{equation}
for $i=1,\ldots,M$ with $\bar{n}_i=0$ for $i<1$ and $i>M$ implied.
In some applications we use a semi-infinite array $({M\to\infty})$ with the second boundary condition replaced by a prescribed (zero or nonzero) limit $\bar{n}_\infty$.
In the following we use the control variables,
\begin{equation}\label{eq:b3} 
\zeta\doteq e^{\beta\mu},\quad \lambda_i\doteq e^{-\beta\mathcal{U}_i},
\end{equation}
with the implication that $\lambda_i=0$ for $i<1$ and $i>M$.

For rods of size $\sigma=1$ the hardcore repulsion does not produce any interference between the $\bar{n}_i$ at different positions.
Equations (\ref{eq:b2}) remain uncoupled.
The density profile $\bar{n}_i$ of rods, which, in this case, coincides with the mass density profile $\rho_i$, reads
\begin{equation}\label{eq:b4} 
\bar{n}_i=\frac{\zeta\lambda_i}{1+\zeta\lambda_i},\quad i=1,\ldots,M.
\end{equation}
Its only structure is that imposed by the potential $\mathcal{U}_i$.

In the case $\sigma=2$ a rod at position $i$ obstructs the placement of a rod positions $i-1$ and $i+1$.
This interference is reflected in Eqs.~(\ref{eq:b2}), which now read
\begin{equation}\label{eq:b5} 
\zeta \lambda_i=\frac{\bar{n}_i(1-\bar{n}_i)}{(1-\bar{n}_i-\bar{n}_{i-1})(1-\bar{n}_i-\bar{n}_{i+1})},\quad i=1,\ldots,M,
\end{equation}
with $\bar{n}_0=\bar{n}_{M+1}=0$ implied.
We solve the coupled Eqs.~(\ref{eq:b5}) by a strategy that also works for $\sigma>2$ as we shall see.
We reduce the set of nonlinear equations into two sets of recursion relations to be solved in sequence.
This method has the benefit of isolating the physical solution.
We begin by introducing the auxiliary quantities,
\begin{equation}\label{eq:b6} 
h_i\doteq\frac{\bar{n}_i}{1-\bar{n}_i-\bar{n}_{i+1}},\quad i=1,\ldots,M.
\end{equation}
We thus convert (\ref{eq:b5}) into
\begin{equation}\label{eq:b7} 
\zeta\lambda_i=h_i+h_ih_{i-1},\quad i=1,\ldots,M,
\end{equation}
from which we determine the $h_i$ recursively:
\begin{equation}\label{eq:b8} 
h_i=\frac{\zeta\lambda_i}{1+h_{i-1}},\quad i=1,\ldots,M.
\end{equation}
The $\bar{n}_i$ are then, in turn, determined recursively via (\ref{eq:b6}):
\begin{equation}\label{eq:b9} 
\bar{n}_i=\frac{h_i}{1+h_i}(1-\bar{n}_{i+1}),\quad i=M,\ldots,1,
\end{equation}
producing the explicit form
\begin{equation}\label{eq:b10} 
\bar{n}_i=\sum_{l=0}^{M-i}(-1)^l\prod_{k=0}^l\frac{h_{i+k}}{1+h_{i+k}},\quad i=1,\ldots,M.
\end{equation}

A different rendition of that solution is derived from  (\ref{eq:b9}) with use of (\ref{eq:b8}):
\begin{equation}\label{eq:b11} 
\bar{n}_i=\frac{h_i}{1+h_i+g_{i+1}},\quad i=1,\ldots,M,
\end{equation}
where the $g_i$ are generated recursively:
\begin{equation}\label{eq:b12} 
g_i=\frac{\zeta\lambda_i}{1+g_{i+1}},\quad i=M,\ldots,1.
\end{equation}

If the external potential is symmetric under under reflection, as in power-law traps or boxes with rigid walls, we have $\lambda_{M+1-i}=\lambda_i$, $i=1,\ldots,M$. 
The density profile of rods must then also exhibit that symmetry.
To make this symmetry transparent we recognize that we have $g_{M+1-i}=h_i$, $i=1,\ldots,M$, under these circumstances.
We can then transform $\bar{n}_{M+1-i}$ into $\bar{n}_i$ as follows:
\begin{widetext}
\begin{align}\label{eq:b13} 
\frac{h_{M+1-i}}{1+h_{M+1-i}+g_{M+2-i}} &= 
\frac{\displaystyle \frac{\zeta\lambda_{M+1-i}}{1+h_{M-i}}}
{\displaystyle 1+ \frac{\zeta\lambda_{M+1-i}}{1+h_{M-i}}+g_{M+2-i}}
=\frac{\zeta\lambda_{M+1-i}}{(1+g_{M+2-i})(1+h_{M-i})+\zeta\lambda_{M+1-i}} \nonumber \\
&=\frac{\zeta\lambda_i}{(1+h_{i-1})(1+g_{i+1})+\zeta\lambda_i}
=\frac{\displaystyle \frac{\zeta\lambda_{i}}{1+h_{i-1}}}
{\displaystyle 1+g_{i+1}+ \frac{\zeta\lambda_{i}}{1+h_{i-1}}}=\frac{h_i}{1+h_i+g_{i+1}}.
\end{align}
\end{widetext}

Now we generalize this method to rods of unrestricted size on the lattice: $\sigma=1,2,\ldots$. 
A rod of size $\sigma$ at position $i$ obstructs the placement of rods at positions $i\pm1,\ldots, i\pm(\sigma-1)$.
The right-hand side of (\ref{eq:b2}) contains $\bar{n}_i$ at $2\sigma-1$ consecutive positions.
The coupled equations now involve polynomials of order $\sigma$.
The boundary conditions require that we set $\bar{n}_i=0$ for $i=-\sigma+2,\ldots,0$ and $i=M+1,\ldots,M+\sigma-1$ in these $M$ coupled equations.
The auxiliary quantities
\begin{equation}\label{eq:b14} 
h_i\doteq\bar{n}_i\left[1-\sum_{k=0}^{\sigma-1}\bar{n}_{i+k}\right]^{-1},\quad i=1,\ldots,M
\end{equation}
then convert (\ref{eq:b2}) into the set of recursion relations
\begin{equation}\label{eq:b15} 
h_{i} = \zeta \lambda_{i}\prod_{k=1}^{\sigma-1}\big(1+h_{i-k}\big)^{-1}, \quad i=1,...,M.
\end{equation}
For given $h_i$ the $\bar{n}_i$ then follow recursively from (\ref{eq:b14}):
\begin{equation}\label{eq:b16} 
\bar{n}_i=\frac{h_i}{1+h_i}\left(1-\sum_{k=1}^{\sigma-1} \bar{n}_{i+k}\right),\quad i=M,\ldots,1.
\end{equation}

%
\section{Pair distribution function}\label{sec:appc}
%
Consider a system of $N$ hard rods of size $\sigma$ at positions $i_k$, $k=1,\ldots,N$ on a 1D lattice of $L$ sites. 
The system is confined by hard walls at positions $i_0$ and $i_{N+1}$. 
The lattice partition function
\begin{align}\label{eq:part_fct}
Z_N(i_0,i_{N+1}) = &\sum_{i_N=i_0+N\sigma}^{i_{N+1}-\sigma}\hspace{1mm}\sum_{i_{N-1}=i_0+(N-1)\sigma}^{i_{N+1}-2\sigma}\ldots \nonumber\\
&\times\sum_{i_1=i_0+\sigma}^{i_{N+1}-N\sigma}e^{-\beta mg\sum_{k=1}^{N}i_k}
\end{align}
with variable change $j_k = i_k - i_0 - k\sigma$
and excess volume defined as
\begin{equation}\label{eq:excess_vol}
L_{ex} = L-N\sigma = i_{N+1}-i_0 - (N+1)\sigma,
\end{equation}
becomes
\begin{align}\label{eq:part_fct_2}
Z_N(i_0,i_{N+1}) = \frac{1}{N!}&e^{-\beta mg(Ni_0+\frac{1}{2}\sigma N(N+1))}\sum_{j_N=0}^{L_{ex}}\sum_{j_{N-1}=0}^{L_{ex}}\ldots\nonumber\\
&\times\sum_{j_1=0}^{L_{ex}}e^{-\beta mg\sum_{k=1}^{N}j_k}
\end{align}
and, after summation over $j_k$,
\begin{align}\label{eq:part_fct_explicit}
Z_N(i_0,i_{N+1}) = \frac{1}{N!}&e^{-\beta mg(Ni_0+\frac{1}{2}\sigma N(N+1))}\nonumber\\
&\times\left(\frac{1-e^{-\beta mg(L_{ex}+1)}}{1-e^{-\beta mg}}\right)^{\!\!N}.
\end{align}
The density and pair distribution functions are then expressible as follows in terms of partial partition functions $Z_k(l,m)$ representing systems of $k$ rods confined by hard walls at positions $l$ and $m$  \cite{Percus:1982,Davis:1990,Ibsen/etal:1997,Buschle/etal:2000a}:
\begin{align}\label{eq:neq_recursion}
\rho(i)=\frac{e^{-\beta mgi}}{Z_N(i_0,i_{N+1})} 
\sum_{k=1}^{N}Z_{k-1}(i_0,i)Z_{N-k}(i,i_{N+1}),
\end{align}
\begin{align}\label{eq:rho2_recursion}
\rho^{(2)}(i,j)=&\frac{e^{-\beta mg(i+j)}}{Z_N(i_0,i_{N+1})}\\
&\times\sum_{k<l=1}^{N}Z_{k-1}(i_0,i)Z_{l-k}(i,j)Z_{N-l}(j,i_{N+1}).\nonumber 
\end{align}
A lattice version of expression (\ref{eq:p-rho2-non-interacting}) for the average microscopic pressure to be used in Sec.~\ref{sec:grav-micro} for comparison with the DFT pressure profiles thus reads
\begin{equation}\label{eq:micscopic_pressure}
\bar{p}_\mathrm{mic}(i) = k_BT\rho(i) + k_BT\sum_{k=0}^{\sigma}\rho^{(2)}(i-k,i+\sigma - k).
\end{equation}

\end{document}